\documentclass[11pt,letter]{article}

\usepackage{graphicx}%

\usepackage{amsmath,amssymb,amsfonts}%
\usepackage{amsthm}%
\usepackage{mathrsfs}%

\usepackage{geometry}

\def\bea {\begin{eqnarray}}
\def\eea {\end{eqnarray}}

\def\be {\begin{equation}}
\def\ee {\end{equation}}

\def \nn {\nonumber}
\def \ep{\epsilon}
\def\P {{{\cal O}_A}}
\def\Ps{{{\cal O}_{A^*}}}
\def\vep {\varepsilon}
\def\As{A^*}
\def\C{{\cal C}}

\begin{document}

\begin{flushright}\mbox{YITP-SB-25-02}\end{flushright}

\renewcommand{\thefigure}{\arabic{figure}}

\begin{center}
{\bf\Large Coordinate light-cone-ordered perturbation theory\footnote{Published online in the European Physical Journal Special Topics on August 4, 2025.}
}
\end{center}
\begin{center}
{Ozan Erdo\u{g}an and George Sterman}\\
\vspace{8pt}
C.N.\ Yang Institute for Theoretical Physics and Department of Physics and Astronomy\\ 
Stony Brook University, Stony Brook, New York, 11794-3840, USA\\
\vspace{8pt}
\today
\end{center}

\begin{abstract}

We review the development of light-cone-ordered perturbation theory in coordinate space (C-LCOPT).
Compared to light-cone-ordered perturbation theory in momentum space (LCOPT), the role of intermediate states in LCOPT is played  in C-LCOPT
by paths, which are ordered sequences of lines and vertices that connect pairs of external points.
Each path denominator of C-LCOPT equals the difference between the separation of the minus coordinates
of the beginning and ending points of the path and the sum of the light-cone distances of all lines along the path computed from their plus and transverse coordinates.
We observe that this method, originally applied to amplitudes, can be extended to cross sections,
which are given in terms of closed paths reminiscent of Schwinger-Keldysh formalisms. 

\end{abstract}

\tableofcontents

\section{Introduction}\label{sec1}

Weak-coupling perturbation theory played an important role in the development of  quantum field theory on
the light cone \cite{Brodsky:1997de}.  Notable early developments include Refs.\ \cite{Bardakci:1968zqb,Chang:1968bh,Kogut:1969xa}.
Choosing variables as light-cone components in momentum and in coordinate space,
\bea
l^\pm\ &=&\ \frac{1}{\sqrt{2}} \left( l^0 \pm l^3 \right ), \nn\\[2mm]
x^\pm\ &=&\ \frac{1}{\sqrt{2}} \left( x^0 \pm x^3 \right ),
\label{eq:lc-variables}
\eea
results in a simplification of perturbative amplitudes, first derived by applying an infinite-momentum boost \cite{Weinberg:1966jm}.
This formalism immediately found use in applications to the then-new parton picture for deep-inelastic scattering \cite{Bjorken:1970ah},
and in describing bound systems at high energies and momentum transfers \cite{Lepage:1980fj}.   As we review shortly, terms in
the resulting diagrammatic expansion have an intuitively appealing correspondence to physical processes.   They also incorporate
unitarity in a transparent fashion.   This made the resulting light-cone-ordered perturbation theory of particular use
in proofs of infrared safety for jet cross sections \cite{Sterman:1978bj} and for weighted cross sections \cite{Sterman:1979uw}.

In this article, we shall primarily review and extend the results of Ref.\ \cite{Erdogan:2017gyf}.
We will return to the general reasoning of Refs.\ \cite{Chang:1968bh} and \cite{Kogut:1969xa}, 
starting with Green functions expressed in coordinate space, 
\bea
G(x_1, \dots x_n) \ = \ \langle 0 | T ( \phi(x_1) \dots \phi(x_n) ) |0 \rangle \, ,
\label{eq:topt}
\eea
applying the light-cone variables
of Eq.\ (\ref{eq:lc-variables}) to the perturbative expansion.  
As we shall see, there are similarities and also important differences in the resulting expressions
compared with momentum space.
Qualitatively, the role of intermediate states, central to light-cone-ordered perturbation theory
in momentum space \cite{Chang:1968bh,Kogut:1969xa}, is taken by ``paths" leading
from earlier to later vertices, ordered in terms of $x^+$ coordinates.
Extending the work of Ref.\ \cite{Erdogan:2017gyf}, we will show how  this formalism
provides a natural expression for inclusive cross sections, in which infrared safety 
appears in a manner that complements its realization directly in momentum space.
As it turns out, this is relatively easy to do for purely massless fields, and we will
concentrate on these.   Most of our presentation is in terms of scalar fields, although
we describe how our results are applicable as well to theories with spin, including gauge theories. 

Our motivation in developing a coordinate analysis of massless perturbation theory
is not so much to provide a new method for calculating Green functions or cross sections.
Rather, as in the use of light-cone-ordered perturbation theory 
to prove the all-orders finiteness of infrared safe observables in momentum space  \cite{Sterman:1978bj,Sterman:1979uw}, it is
to provide an additional perspective on the infrared behavior of quantum fields.   
In particular, we believe that it may be useful to provide a space-time
picture for the cancellations of infrared divergences.   In this context, the
largest-time equation  \cite{Veltman:1963th,tHooft:1973wag} will play an important role.

Some of the results presented below are to our knowledge new to the literature.
The ordering of vertices along the coordinate light cone
and its connection to cancellations in inclusive cross sections, however, are  familiar from treatments
of radiation by charged particles in media \cite{Baier:1996sk,Zakharov:1997uu,Zakharov:1998sv,Arnold:2023qwi,Caucal:2023fsf}, and the related evolution of particle and nuclear cross sections
at very high energies \cite{Balitsky:1995ub,Kovchegov:1999yj,Gelis:2010nm}.
Even more closely related is
``flow-ordered" perturbation theory,
developed 
 in Refs.\ \cite{Borinsky:2022msp} and \cite{Salas-Bernardez:2023aqt} (see also Ref.\ \cite{Salas-Bernardez:2023zzv}), where 
a new formalism has been developed with many features in common with the light-cone treatment presented here. In \cite{Borinsky:2022msp}, the role of  minus
light-cone integrations here is taken by time integrations, so that the comparison is
analogous to the comparison of LCOPT to time-ordered perturbation theory in momentum space.
Compared with time ordered perturbation theory, the flow ordered coordinate approach has the advantage that it provides expressions with fewer diagrams
than would expected from time ordered perturbation theory in momentum space.   This reduction in diagrams is related to those
found in momentum space in Refs. \cite{Capatti:2022mly} and \cite{Sterman:2023xdj}.

To set the stage, we  briefly review the perturbative method
in momentum space (LCOPT) introduced in Refs.\ \cite{Chang:1968bh,Kogut:1969xa},
specialized to  (massive or massless) scalar fields.  
  We represent an arbitrary (truncated) diagram with external lines of momenta $p_i$, $N$ internal vertices, $L$ lines and ${\cal L}$ loops, as
\bea
G(\{ p_a\}) \ = \  {(-ig)^N}\; \int\prod_{\mathrm{loops}\, i=1}^{\cal L} \frac{d^4l_i}{(2\pi)^4} \prod_{\mathrm{lines}\, j=1}^L \frac{i}{k^2_j(l_i,p_a) -m^2 + i\ep} \ .
\label{eq:G-mtm} 
\eea
 We suppress symmetry factors and other overall constants.
The inclusion of spin, as in a gauge theory, would require modified numerator factors in this expression, which, although important for
a full treatment \cite{Chang:1968bh,Kogut:1969xa,Brodsky:1997de}, do not modify the results we are after.  The procedure leading to 
light-cone-ordered perturbation theory can be summarized as:  

\begin{enumerate}

\item Rewrite the product of loop minus momentum integrals as a product of line
minus momentum integrals by introducing minus conservation delta functions at each vertex. \\

\item  Represent the delta functions as integrals
over  conjugate light-cone plus coordinates.   \\

\item  Write the product of plus coordinate integrals as a sum over ordered sectors.   \\

\item  In each sector perform the now-trivial contour integrals for line minus momenta at fixed vertex order, determined by the mass shell pole at
$k_j^- = (k_{j \perp}^2+m^2 -i\ep)/2k_j^+$ for all $j=1,\dots L$, and finally, \\

\item  Perform the plus coordinate integrals.   

\end{enumerate}

This procedure, which we will use as the inspiration for our treatment of coordinate integrals below, leaves us with an expression for
$G(\{ p_a\})$ as a sum over ordered diagrams,
\bea
 G(\{ p_a\}) \ = \sum_{\cal O} G_{\cal O} (\{ p_a\})\, ,
 \label{eq:G-sum-mtm}
 \eea
 where ${\cal O}$ represents an ordering of vertices, and $G_{\cal O}$ an expression in which loops have three integrals over momentum components to carry out, plus  and transverse.
For a scalar diagram with lines of mass $m$, each such ordered diagram is of the form \cite{Chang:1968bh,Kogut:1969xa},
  \bea
-i\, G_{\cal O}(\{ p_a\})\ & = & \ g^N\;  \prod_{{\rm loops}\ i} \int\,\frac{dq^+_id^2q_{i\perp}}{(2\pi)^3}\, \prod_{{\rm lines}\ j} \frac{\theta(k_j^+)}{2k_j^+}
\nn\\[2mm]
&\ & \hspace{10mm} \times
 \prod_{{\rm states}\ s=1}^{N-1}\   \frac{1}{ P_{\rm ext}^-{}{(s)}\ - \sum_{k_j\in s} \frac{k_{j,\perp}^2+m^2}{2k_j^+}\ +\ i\ep }\, ,
 \label{eq:lcopt-mtm-basic}
\eea
with $N$ the number of interaction vertices, with coupling $-ig$, which corresponds to $N-1$ intermediate states.   The line momenta, denoted by $k_j$, are linear combinations of loop and external momenta.
The denominators in this expression are the result of the final integrals over plus coordinates, and have the interpretation of ``light-cone energy deficits" for intermediate states of the scattering process, the states consisting of the sets of line $s$ between successive vertices in the plus ordering.  This deficit is simply the difference between  $P_{\rm ext}^-{}{(s)}$, the total external minus momentum flowing into the diagram before state $s$, and the on-shell minus momenta of all lines in state $s$.  When denominators are all real and nonzero, the overal factor of $-i$ on the left-hand side corresponds to a real contribution to the $T$ matrix.  

In the next section, we shall follow Ref.\ \cite{Erdogan:2017gyf} and carry out an analogous evaluation of Green functions in coordinate space, with the aim of doing one integral at each vertex.  Unlike the momentum analysis we have just outlined, however, we restrict ourselves to massless fields.   While extensions to massive fields are possible, our interest in the infrared behavior makes this restriction natural as a starting point.

\section{Coordinate Integrals for Scalar Amplitudes: Paths}
\label{sec2}

\subsection{Light-cone integrals for massless fields}

We now turn to  the path formalism for coordinate-space Green functions given in Ref.~\cite{Erdogan:2017gyf}, before proceeding with its application to cross sections\footnote{With applications in mind, we often refer to these diagrammatic contributions as amplitudes below.}. 
The coordinate representation of a generic Green function corresponding to a connected scalar diagram with  external points, $x_a$, can be expressed as an integral over the positions of $N$ internal vertices, $y_i$, of the product of propagators for $L$ lines.
Lines, labelled by index $j$, connect vertices separated by vectors $z^\mu_j$.   In these terms, diagrams, $G(\{x_a\})$, are given by
\be
G(\{ x_a\}) \ = \  \frac{(-ig)^N}{(4\pi^2)^L}\; \prod_{\mathrm{vertices}\,  i=1}^N \int\, d^4y_i \prod_{\mathrm{lines}\, j=1}^L \frac{1}{-z^2_j(y_i,x_a) + i\ep} \, ,
\label{eq:G-massless} 
\ee
in terms of massless coordinate space propagators,
\bea
\Delta(z) &=& \int \frac{d^4p}{(2\pi)^4} \, e^{-ip\cdot z}\, \frac{i} {p^2+i\ep }
\nn\\[2mm]
&=& \frac{1}{4\pi^2}\, \frac{1}{ -z^2 + i\ep }\, .
\label{eq:causal-propagator}
\eea
In Eq.\ (\ref{eq:G-massless}), the arguments of the propagators, $z_j$, for lines $j$ are given as linear combinations of the vertex positions, $y_i$ for internal vertices and $x_a$ for external,
\bea
z^\mu_j \ =\ \eta_{ji}y^\mu_i + \eta'_{ja}x^\mu_a  \, ,
\label{eq:zs-from-ys}
\eea
with $\eta_{ji}$ and $\eta'_{ja}$ incidence matrices whose elements take on the values $\pm 1$ or $0$.
Following a slight variation of the steps given in detail by \cite{Erdogan:2017gyf}, we work by analogy to the momentum-space analysis sketched above.  

We begin here with an integral representation \cite{Borinsky:2022msp} for the massless causal propagators of Eq.\ (\ref{eq:G-massless}),
\bea
\frac{1}{-z^2_j(y_i,x_a) + i\ep} = -i\, \int_{-\infty}^{\infty} dE_j^+\, \frac{\theta\left( E_j^+z_j^+ \right)}{2|z_j^+|}\,
 e^{ -iE_j^+ \left(  \eta_{ji}y^-_i + \eta'_{ja}x^-_a - \frac{z_{j \perp}^2+i\ep}{2z_j^+} \right ) }
  \, ,
\nn\\
\label{eq:amp-rep}
\eea
where we have exhibited the explicit definition of $z_j^-$ from Eq.\ (\ref{eq:zs-from-ys}).
Notice the step function that links the signs of $E_j^+$, a quantity that we will refer to below as ``light-cone energy", and component $z_j^+$.  
The $E_j^+$ integral extends from zero to infinity when $z_j^+$ is positive, and from minus infinity to zero when $z_j^+$ is negative.

\subsection{Vertex ordering}

We now make use of the role played by plus coordinates of lines in the representation, Eq.\ (\ref{eq:amp-rep}), by ordering vertices by the values of their plus momenta, $y_i^+$, writing the original integral as a sum of these orderings, here labeled ${\cal O}$,
\bea
 G(\{ x_a\}) \ = \sum_{\cal O} G_{\cal O} (\{ x_a\})\, .
 \label{eq:G-sum-coord}
 \eea
 The orderings, $\cal O$, define ``partially ordered sets" or ``posets", in which each vertex is only ordered with respect to those vertices to which it connects directly by one or more lines \cite{Erdogan:2017gyf}.\footnote{This is in distinction to the full orderings of vertices in LCOPT.   See also Ref.\ \cite{Sterman:2023xdj}.}   That is, if $z^+_j = y^+_k - y^+_l>0$ then $y_k>y_l$.   This partial ordering, which is taken to be transitive, is fully determined by the signs of the $z_j^+$, and is all we will need.
 
  Next, substituting Eq.\ (\ref{eq:amp-rep}) for the propagators in the amplitude, Eq.\ (\ref{eq:G-massless}), in any of the regions specified by ${\cal O}$, all dependence on the $y_i^-$ coordinates is in the exponents.  
 Integrating over the minus components of the internal vertices gives a light-cone energy-conservation delta function for each internal vertex, 
\bea 
G_{\cal O}\left (\{ x_a\} \right ) 
&  =  &  
 \frac{(-ig)^N}{(4\pi^2)^L}\; (-i)^L\, \prod_{i\in N}  \int d^2y_{i\perp} \int_{\cal O} dy_i^+\, \int_{-\infty}^\infty dy_i^-
 \nn\\[2mm]
 &\ & \times \prod_{j\in L} \int_{-\infty}^\infty dE^+_j\,  \frac{ \theta \left (z^+_jE^+_j \right ) }{2|z^+_j|}
 e^{-iE^+_j(\eta_{ji}y^-_i + \eta'_{ja}x^-_a)}\,
  e^{iE^+_j\frac{z^2_{j\perp}+i\ep}{2z^+_j}}  \nonumber \\[2mm]
 &\ & \hspace{-20mm} =\
  \frac{(-ig)^N\, (-i)^L}{(2\pi)^{2L-N}} \, \prod_{i\in N} \int d^2y_{i\perp} \int_{\cal O} dy_i^+   \nn  \\[2mm] 
 & &  \times \prod_{j\in L} \int_{-\infty}^\infty dE^+_j\, \, \frac{\theta \left (z^+_jE^+_j \right )}{2|z^+_j|}\, 
 e^{-iE^+_j \left (\eta'_{ja}x^-_a-\frac{z^2_{j\perp}+i\ep}{2z^+_j} \right ) }
 \, \prod_{i\in N}\delta \left (E^+_j\eta_{ji} \right ) \, . \nn\\
 \label{eq:E-integrals-1}
 \eea
 The step functions and delta functions in this expression limit the set of vertex orderings, ${\cal O}$ that can contribute to the diagram.  Here and below, we use $L$ to represent the set of lines as well as the number of elements in the set.
 
As noted above, the step functions require that light-cone energies flow in the same direction as the plus component of the line separation.  That is, when $z^\mu= y_k^\mu - y_l^\mu$, component $y^+_k-y^+_l$ is positive (negative) when $E_j^+$ is positive (negative).  Thus, the product of step functions vanishes {\it except} for orderings of the $y_i^+$ such that positive light-cone energies, $E_j^+$ flow from vertices with smaller values of the plus components to vertices with larger values, throughout the diagram.   

The delta function associated with each internal vertex in Eq.\ (\ref{eq:E-integrals-1}) sets $\sum_j E^+_j\eta_{ji}=0$, which ensures that light-cone energy is conserved at each internal vertex.    That is, each internal vertex must receive positive light-cone energy from at least one earlier vertex, and must provide positive light-cone energy to at least one later vertex.  This eliminates, as in the case of momentum space LCOPT, vertices at which particles emerge from the vacuum or are absorbed by it.\footnote{This discussion does not apply to vacuum diagrams, whose convergence properties make the choice of  light-cone coordinates nontrivial in momentum as well as coordinate space, as noticed early in Ref.\ \cite{Chang:1968bh} and \cite{Yan:1973qg}. Further developments are reviewed in \cite{Collins:2018aqt}.}    We label vertex orderings that satisfy the requirements of both the step and delta functions by subscript $\P$, which make up a subset of the ${\cal O}$.\footnote{We may think of subscript $A$ as standing for ``allowed" or for ``amplitude".   We will use $A^*$ in this context for allowed orderings in complex conjugate amplitudes below.}

Our considerations on ordering so far apply to the $N$ vertices whose positions are integrated, the internal vertices, defined by the interaction Lagrange density of the theory.   For this discussion, we take the positions of the remaining, external, vertices, $\{x_a\}$, as fixed.   In the simplest case, as in the expansion of a time-ordered product of fields like Eq.\ (\ref{eq:topt}), each external vertex is connected to an internal vertex by a single line.   For a given ordering of the internal vertices, we can classify such external vertices as ``incoming" or ``outgoing".   For external vertices connected by one or more lines,
 we identify as incoming those that are connected only to vertices with larger $y_i^+$.   Similarly, outgoing external vertices connect only to internal vertices with smaller $y_i^+$.  Any external vertex that is connected to the remainder of the diagram by only a single line is clearly either incoming or outgoing.   
 
 In summary,  for the diagrams we consider, any allowed ordering $\P$  has a unique set of external vertices $\{x_a\}_{\rm in}$ that are incoming and a unique set of external vertices $\{x_b\}_{\rm out}$ that are outgoing.  The positions of external vertices are not integrated, so that energy can flow into and out of the diagrams through its external vertices.   Note that we can (and will below) set the positions of two incoming vertices $x_a$ to be equal, and treat this ``composite" vertex as incoming.   We denote any nonvanishing contribution in the sum over orders as $G_{\P}\left (\{x_b\}_{\rm out},\{x_a\}_{\rm in}\right )$, so that Eq.\ (\ref{eq:G-sum-coord}) becomes
  \bea
  G\left (\{x_b\}_{\rm out},\{x_a\}_{\rm in}\right )
 &=&
 \sum_\P  
 G_{\P}\left (\{x_b\}_{\rm out},\{x_a\}_{\rm in}\right )\, .
 \label{eq:G-GOA}
 \eea
 For any nonvanishing $G_\P$, the light-cone energy $E_j^+$ of any line, $j$, is integrated over positive or negative values, but not both.    It is positive when the corresponding $z_j$ is pointed forward  in this plus component ordering, and negative when it is backward.   For a given fixed ordering, however, we can always choose to reverse the signs of the two incidence matrix elements, $\eta_{ji}$ and $\eta'_{ja}$, for all lines with $z_j^+<0$, which then reverses the signs of the corresponding light-cone energies, making them positive for these values of $j$.  We denote these modified incidence matrix as $\eta_{ji}^{(\P)}$ and ${\eta'}^{(\P)}_{ja}$. With such a change of notation, all $E_j^+$ and $z_j^+$ can separately be taken as positive, and all allowed orders  $G_{\P}$ involve only integrals of $E_j^+$ from zero to infinity,
 \bea
  G_{\P}\left (\{x_b\}_{\rm out},\{x_a\}_{\rm in}\right )
 &=& 
 \frac{(-ig)^N\, (-i)^L}{(2\pi)^{2L-N}} \, \prod_{i\in N} \int d^2y_{i\perp} \int_{\P} dy_i^+   \nn  
 \\[2mm] 
 & & \hspace{-10mm}  \times \prod_{j\in L} \frac{\theta \left (z^+_j \right )}{2 z^+_j}\,  \int_0^\infty dE^+_j\, \, 
 e^{-iE^+_j \left ({\eta'}^{(\P)}_{jc} x^-_c-\frac{z^2_{j\perp}+i\ep}{2z^+_j} \right ) }
 \, \prod_{i\in N}\delta \left (E^+_j\eta_{ji}^{(\P)} \right ) \, . \nn\\
 \label{eq:E-integrals-positive}
\eea
Note that now the sum over external vertices in the exponent is through index $c$, which goes over both incoming ($a$) and outgoing ($b$) external vertices.   We emphasize that in this form, incidence matrices have been adjusted, as indicated by superscript $({\cal O}_A)$, without changing the values of the integrals.   We now turn to the light-cone energy integrals.

\subsection{Light-cone energy integrals}

We are now ready to carry out all the $L$ light-cone energy integrals in Eq.\ (\ref{eq:E-integrals-positive}), using the delta functions for any allowed order $\P$.   For a diagram with $E$ external and $N$ internal vertices, the number of integrals left by the delta functions is $L-N$, the number of lines minus the number of internal vertices.   This is  $ E + {\cal L} -1$, the number of external vertices plus the number of loops, ${\cal L}$, in the diagram minus one.   We will construct the remaining integrals to have a particular graphical intepretation, as energies flowing along paths between external vertices, covering the entire allowed space of line light-cone energies.   

We will see that this choice of integration requires in general a sum over terms, with different choices of independent paths.   For a general diagram, more than one such ``covering set" of paths is necessary to fill the entire light-cone energy integral of a given term $G_\P$ in Eq. (\ref{eq:E-integrals-positive}).  
Labeling a complete set of such covering paths for the integrals of $G_\P$ by $\C[\P]$, we will show that the full allowed vertex order is given as a sum \cite{Erdogan:2017gyf},
\bea
G_{\P}\left (\{x_b\}_{\rm out},\{x_a\}_{\rm in}\right )
\ =\ \sum_{\C \in \C[\P]}
G^{(\C)}_{\P}\left (\{x_b\}_{\rm out},\{x_a\}_{\rm in}\right )\, ,
\label{eq:GOtoGCO}
 \eea
 where we will be able to carry out the light-cone energy integrals in each  $G_\P^{(\C)}$.   
 
 Each term,  $G_\P^{(\C)}$ in Eq.\ (\ref{eq:GOtoGCO}) will be given as an integral over  $E + {\cal L} -1$  independent path light-cone energies, ${\cal E}_\alpha^{(\C)}$, which flow from incoming to outgoing vertices.  The light-cone energy of every line in the diagram is a sum of these path energies,
 \bea
 E_j^+\ =\ \sum_{{\rm paths}\  \alpha=1}^{L-N} \sigma^{({\cal C})}_{j\alpha} {\cal E}_\alpha^{(\C)}\, ,
 \label{eq:Ej-Ealpha}
 \eea
 with coefficient $\sigma^{(\C)}_{j\alpha}=1$ if path $\alpha$ passes through line $j$, and zero otherwise.   Here all path light-cone energies, ${\cal E}_\alpha^{(\C)}$, and hence all line energies $E_j^+$, are integrated  from zero to infinity.   As we shall see, however, for many diagrams a single incidence matrix does not define a full covering of the allowed integration volume in light-cone energy space.
Let us denote the complete line energy integration volume  of $G_\P$ by $R[G_\P]$, and the corresponding path energy integration volume of $G^{({\cal C})}_\P$  by $R[G^{(\C)}_\P]$.   Clearly, using (\ref{eq:Ej-Ealpha}), every point in the path energy space corresponds to a unique point in line energy space.  In the next subsection we will show, and for now we will assume, that we can construct the coverings $\C[\P]$ in such a way that the volumes of the coverings fill the original volume,  
 \bea
 R[G_\P]\ = \cup_{\C \in \C[\P] }\ R[G^{(\C)}_\P]\, ,
  \label{eq:GC-fills}
 \eea
 and that they are disjoint, up to lower-dimensional volumes,
 \bea
 {\cal C}'\ne {\cal C }\ \Rightarrow \ R[G^{(\C')}_\P]\ \cap \ R[G^{(\C)}_\P] \ =\ \phi\, .
 \label{eq:GC-disjoint}
 \eea
 To do the light-cone energy integrals of Eq.\ (\ref{eq:E-integrals-positive}) we can therefore make the replacement,
  \bea
 \prod_{j\in L} \int_0^\infty dE^+_j
 \, \prod_{i\in N}\delta \left (E^+_j\eta_{ji}^{(\P)} \right ) \
 \rightarrow 
 \sum_{\C \in {\cal C}[\P]}\, \prod_{\alpha =1}^{L-N}\, \int_0^\infty d{\cal E}^{(\C)}_\alpha\, .
 \label{eq:volumes}
 \eea
 Each term on the right of this relation defines the integration region for term $G_\P^{(\C)}$ on the right of Eq.\ (\ref{eq:GOtoGCO}).
 This relation is similar to the familiar replacement of line momenta by loop momenta in momentum space.  
 Like the choice of loop momenta,  the choice of covering paths is not unique, although we chose them always to appear in sums with unit coefficients.
 In the next section, we will describe a method for generating consistent choices of the incidence matrices $\sigma_{j\alpha}^{(\C)}$ in
  Eq.\ (\ref{eq:Ej-Ealpha}), which manifestly satisfy the conditions of Eqs.\ (\ref{eq:GC-fills}) and (\ref{eq:GC-disjoint}),
 and hence define the terms $G_\P^{(\C)}$ in Eq.\ (\ref{eq:GOtoGCO}).
 
 For now, substituting (\ref{eq:volumes}) into the original line integral form of Eq.\  (\ref{eq:E-integrals-positive}), and using Eq.\ (\ref{eq:Ej-Ealpha}) to eliminate line light-cone energies in favor of path light-cone energies, we are left with 
 \bea
   G^{(\C)}_{\P}\left (\{x_b\}_{\rm out},\{x_a\}_{\rm in}\right )
 &=& 
 \frac{(-ig)^N\, (-i)^L}{(2\pi)^{2L-N}} \, \prod_{i\in N} \int d^2y_{i\perp} \int_{\P} dy_i^+ \, 
  \nn  \\[2mm] 
 & & \hspace{-10mm} \times\, \prod_{j\in L} \,  \frac{\theta(z_j^+)}{2 z^+_j} \ \prod_{\alpha \in L-N} \int_0^\infty d{\cal E}^{(\C)}_\alpha\, \,  
 e^{-i{\sigma^{(\C)}_{j\alpha}\cal E}^{(\C)}_\alpha \left ({\eta'}_{jc}^{({\P})}x^-_c-\frac{z^2_{j\perp}+i\ep}{2z^+_j} \right ) }
 \, . 
 \nn\\
 \label{eq:E-integrals-positive-path}
\eea
 The integrals for the path light-cone energies are now simple, and we derive
  \bea 
G^{(\C)}_{\P}\left (\{x_b\}_{\rm out},\{x_a\}_{\rm in}\right ) \ &=&\
  (2\pi )^{N-2L}\, (-g)^N  \prod_{i\in N} \int d^2y_{i\perp} \int_\P dy_i^+  \prod_{j \in L} \frac{\theta(z_j^+)}{2z_j^+} 
  \nn\\[2mm]
  &\ & \times\
 \prod_{\alpha \in L-N}\  \frac{-1}{x^-_{b_\alpha}\ -\ x^-_{a_\alpha}\ -\ D^{(\C)}_\alpha -i\ep }\, .
 \label{eq:1st-result-massless}
\eea
Here, each path, $\alpha$, is represented by a denominator, with  $x_{b_\alpha}$ the external vertex at which the path ends (with ${\eta'}^{(\P)}_{b\alpha}=1$), $x_{a_\alpha}$ the external vertex where it begins (with ${\eta'}^{(\P)}_{a\alpha}=-1$),   and
$D^{(\C)}_\alpha$ the sum of light-cone minus distances of the lines between the external vertices of path $\alpha$,
\bea
D^{(\C)}_\alpha\ =\ \sum_j \sigma_{j\alpha}^{(\C)}\, \frac{z_{j\perp}^2}{2z_j^+}  \ \equiv \ \sum_{j \in \pi_\alpha}\, \frac{z_{j\perp}^2}{2z_j^+} \, .
\eea
 In the second form, we have introduced notation for paths $\alpha$, as the sets of lines, $j$, for which $\sigma^{(\C)}_{j\alpha}=1$,
\bea
\pi_\alpha\ = \ \left \{ j \, |\, \sigma^{(\C)}_{j\alpha}=1 \right \}\,  .
\label{eq:path-def}
\eea
The path light-cone energies ${\cal E}_\alpha^{(C)}$ run along these lines, from incoming to outgoing vertices.
Equation (\ref{eq:1st-result-massless}) is our basic result \cite{Erdogan:2017gyf}, which, as promised, is a product of denominators that are the differences between the separations of the minus components of external points and the sum of light-cone minus distances defined by the plus and transverse components of the lines connecting the vertices along the path.   
Comparing to Eq.\ (\ref{eq:lcopt-mtm-basic}) for LCOPT, we see the similarity between the roles of energy deficits there, and path differences here.
We now return to the construction of covering sets of paths for a general ordered diagram.

\subsection{Constructing sets of covering paths}
 
 As we have seen, each term $G_\P^{(\C)}$ in  Eq.\ (\ref{eq:GOtoGCO}) corresponds to a set of paths, $\C = \{\pi_\alpha\}$.   Each path, $\pi_\alpha$ extends from an incoming vertex, $a_\alpha$ to an outgoing vertex, $b_\alpha$, always connecting vertices with increasing values of plus coordinate $y_i^+$.  The paths are distinct, and every line and vertex appears on at least one path.  In the presence of loops, there may be more than one path between two external vertices.\footnote{An equivalent but distinct construction is given in Ref.\ \cite{Erdogan:2017gyf}, which  starts with tree diagrams of arbitrary order, then adds loops.   Here, we have opted for an inductive construction, treating the ordered vertices one at a time.}
 
We now describe an inductive construction of the paths, starting from incoming vertices in a diagram ordered in its plus vertex coordinates.  
At each subsequent vertex, we eliminate one additional line light-cone energy integration in favor of the path light-cone energy integrations, with the number of integrals always decreasing by one. We provide here an algorithm for a theory with only three-point vertices.   The extension to higher-point interactions is straightforward.

We start the procedure with the line or lines emerging from the earliest (in plus coordinate) incoming vertex, identifying their light-cone energies as (potential) path light-cone energies.  
We then assume that all lines that emerge from vertices earlier (in $y_i^+$) than an arbitrary fixed vertex have already been expressed as sums of path light-cone energies, as in Eq.\ (\ref{eq:Ej-Ealpha}), and show how to eliminate line energies in terms of path energies for the line or lines that emerge from this vertex.

 \begin{figure}
\centering
\includegraphics[height=5cm]{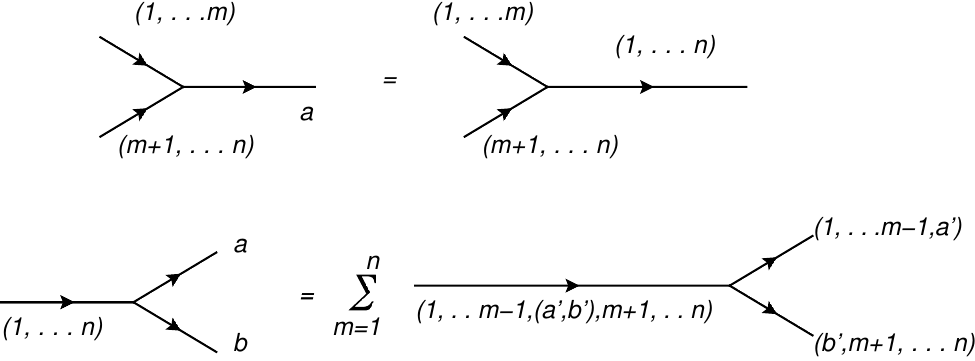} 
\vspace{1cm}
\caption{Algorithm for the construction of paths in a theory with three-point vertices, proceeding from vertices with lower plus components to greater in the amplitude, Eq.\ (\ref{eq:G-massless}).  In the first relation the paths arriving at the vertex are merged.  In the second relation, the set of light-cone energies and paths arriving at the vertex are divided between the two outgoing lines, according to Eq.\ (\ref{eq:Ea-region}).  We sum over the choice  of $m$, the incoming path light-cone momenta that is split.    }  
\label{fig:merge-split}
\end{figure}

\begin{figure}
\centering
\includegraphics[height=6cm]{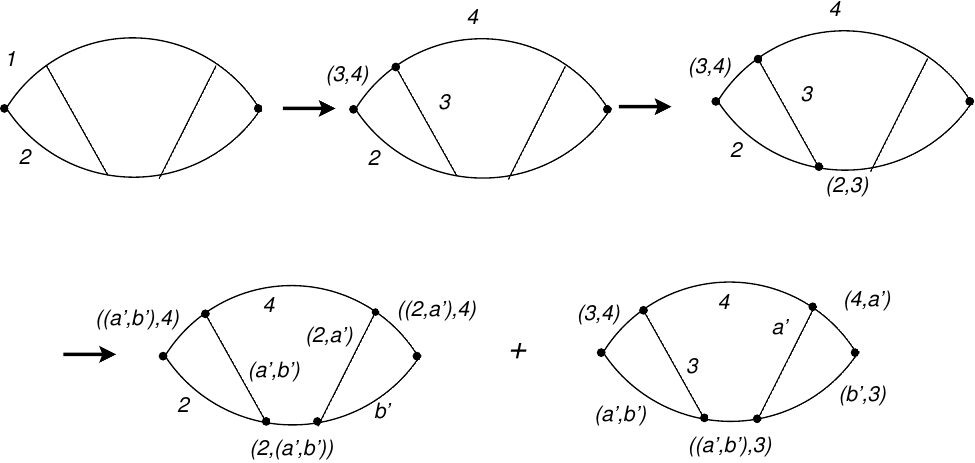} 
\vspace{1cm}
\caption{Example of the construction of paths through an ordered 3-loop diagram using the steps of Fig.\ \ref{fig:merge-split}.  The first step, starting at the earliest external vertex,  identifies provisional path momenta as line momenta ${\cal E}_1$ and ${\cal E}_2$.  Next, at the vertex that follows, ${\cal E}_1$ splits and is replaced by two paths, ${\cal E}_3$ and ${\cal E}_4$.  The notation $(3,4)$ indicates that ${\cal E}_3+{\cal E}_4$ flows on the corresponding line.  At the third step, ${\cal E}_3$ and ${\cal E}_2$ merge.  Next, the line $(2,3)$, with energy ${\cal E}_2+{\cal E}_3$, splits, leading to two terms, in which either ${\cal E}_3$ or ${\cal E}_2$ is replaced by ${\cal E}_{a'}+{\cal E}_{b'}$.  In both cases, the final step is a merge, which leaves the number of terms unchanged. }  
\label{fig:path-example}
\end{figure}

In a theory with only three-point vertices, each vertex of an ordered diagram describes either the merging of two lines into one, or the splitting of one line into two.

Consider first the merging of two lines, illustrated in the upper relation in Fig.\ \ref{fig:merge-split}.  
  The indices $(1, \dots m)$ and $(m+1, \dots n)$ on the lines entering from the left represent the sums of all path light-cone energies that flow through these lines.  When the corresponding propagators merge at the next vertex, the energy integral of the line on the right is simply replaced by the sum of the path energies arriving at the vertex from the left.   The number of line integrals has decreased by one, with the outgoing line carrying the sum of incoming light-cone energies, denoted by $(1, \dots n)$.  
  
   When a line splits, as in the lower relation of Fig.\ \ref{fig:merge-split}, we need a slightly more elaborate rule, which preserves the positivity of all path momenta while covering the full integration region.    
 As above, the parentheses indicate that light-cone energies ${\cal E}^{({\cal C})}_1$ to ${\cal E}^{({\cal C})}_n$ are summed for the incoming line.  Each of the distinguishable paths $1, \dots n$ extends from an incoming vertex to the vertex in question.  On the left of this relation, indices $a$ and $b$ represent the light-cone energies of the corresponding lines that emerge from the vertex, whose sum must equal the total light-cone energy flowing into the vertex from the left, $(1,\dots n)$.    When more than one path energy arrives at the vertex ($n>1$ here), we must choose how these paths flow into the lines labelled $a$ and $b$.   
 
 We begin with an arbitrary ordering of the incoming path light-cone energies, ${\cal E}^{(\C)}_1 \dots {\cal E}^{(\C)}_n$ on the left of the figure.
 We then choose to specify the two light-cone line energies $E_a^+$ and $E_b^+$ according to a set of regions, defined by
  \bea
 \sum_{\alpha=1}^{m} {\cal E}^{({\cal C})}_\alpha\ \ge E^+_a\ \ge \sum_{\alpha=1}^{m-1} {\cal E}^{({\cal C})}_\alpha\, .
 \label{eq:Ea-region}
 \eea
 Each of these inequalities is disjoint, while their union covers the full integration region.   In this way, the construction satisfies the conditions given above in Eqs.\ (\ref{eq:GC-disjoint}) and (\ref{eq:GC-fills}), respectively.
 Applying the momentum conservation delta function at the vertex in each such region, we construct a term in which paths $\alpha=1, \dots m-1$  (and their light-cone energies) flow into line $a$.   Path $m$, with $0<m \le n$, is split into two paths, and we define ${\cal E}^{({\cal C})}_m={\cal E}^{({\cal C})}_{a'}+{\cal E}^{({\cal C})}_{b'}$, where path $a'$ flows into line $a$ and path $b'$ into line $b$.   Finally, the remaining paths ($\alpha \ge m+1$) flow into $b$. We have increased the number of path integrals by one, by replacing ${\cal E}^{(C)}_m$ with ${\cal E}^{(C)}_{a'}$ and ${\cal E}^{(C)}_{b'}$,  but we have decreased the number of line integrals by two, eliminating $E^+_a$ and $E^+_b$, so that overall there is one fewer integral.  
 With this choice of variables and paths, light-cone energies are conserved, and the full integration volume is covered by non overlapping subregions.   
 
 An example at three loops, with one composite external incoming vertex and one outgoing, is given in Fig.\ \ref{fig:path-example}, which illustrates our procedure. We find in this case two sets of paths, as shown in the second line of the figure.  Each set is an element of ${\cal C}[\P]$  in Eq.\ (\ref{eq:GOtoGCO}), with a distinct assignment of paths between the incoming and outgoing vertices, as in Eq.\ (\ref{eq:1st-result-massless}).
 These two elements of the set ${\cal C}[\P]$ have different incidence matrices, $\sigma^{(\C)}_{j\alpha}$, which can be read off from the figure.

\subsection{Example}

\begin{figure}
\centering
\includegraphics[height=3cm]{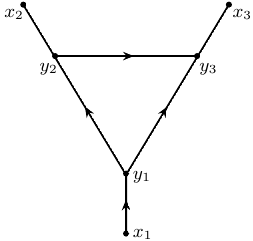} \quad \hspace{1cm} \includegraphics[height=3cm]{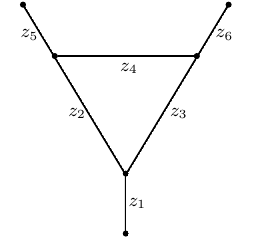}\\
\vspace{3mm}
(a) \hspace{3.75cm} (b)
\vspace{.75cm}
\caption{Labelling of (a) vertices and (b) lines for the example diagram in coordinate space.}  
\label{fig:triangle-example}
\end{figure}

\begin{figure}
\centering
\includegraphics[height=3cm]{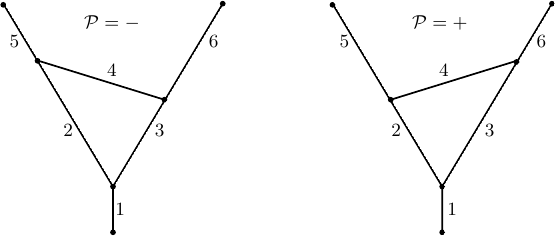}\\
\vspace{3mm}
(a) \hspace{3.75cm} (b)
\vspace{.75cm}
\caption{Vertex light-cone plus coordinate-ordered diagrams discussed in text, given by Eqs.\ (\ref{eq:path-form-example-5minus}) (a) and (\ref{eq:path-form-example-5plus}) (b).}  
\label{fig:ordered-scalar-tri}
\end{figure}

We now illustrate the foregoing general considerations explicitly, with the one-loop example shown in Figs.\ \ref{fig:triangle-example} and \ref{fig:ordered-scalar-tri}.   The diagram shown in these figures, with its external lines, has six vertices, and hence $6!$ orderings.   Restricting ourselves to one incoming ($x_1$) and two
outgoing $(x_{2,3})$ vertices in Fig.\ \ref{fig:triangle-example}, the ordering requirements reduce the number of terms to two, shown in Fig.\ \ref{fig:ordered-scalar-tri}, which differ by the relative orders of the plus coordinates of vertices $y_2^\mu$ and $y_3^\mu$.

In the notation of Eq.\ (\ref{eq:1st-result-massless}), the diagram in Fig.\ \ref{fig:ordered-scalar-tri}a is given by 
\bea
G_{{\cal O}_a}^{(\C_a)} \left (\{x_2,x_3\}_{\rm out},\{x_1\}_{\rm in}\right ) &=& -\,  \frac{(-g)^3}{(2\pi)^9}\,  \prod_{i=1}^3 \int d^2y_{i\perp} \int dy_i^+  \prod_{j=1}^6 \frac{ \theta(z_j^+)}{2z^+_j}  
\nn\\
&\ & \hspace{-36mm} \times\frac{1}{ x_3^--x_1^-  - D_{631}
 -i\ep } \, 
\frac{1}{ x_2^--x_1^- -  D_{521}
-i\ep } \, \frac{1}{ x_2^--x_1^-  -  D_{5431}
-i\ep} \, ,
\label{eq:path-form-example-5minus}
\eea
with its sums of ``light-cone minus coordinates" in the denominators.  These correspond to the three paths in light-cone energy that generate the denominator in this case,
\bea
 D_{i_n,\dots , i_1} = \sum_{j \in \{i_n,\dots i_1\}}     \frac{z^2_{j\perp}+i\ep}{2z^+_{j}} \ .
  \label{eq:Dlines}
  \eea
Similarly, Fig.\ \ref{fig:ordered-scalar-tri}b is given by
\bea
G_{{\cal O}_b}^{(\C_b)} \left (\{x_2,x_3\}_{\rm out},\{x_1\}_{\rm in}\right ) &=& -\,   \frac{(-g)^3}{(2\pi)^9}\,  \prod_{i=1}^3 \int  d^2y_{i\perp} \int dy_i^+  \prod_{j=1}^6 \frac{ \theta(z_j^+)}{2z^+_j}   \nn \\
& \ & 
\hspace{-36mm}\times\frac{1}{ x_2^--x_1^- -  D_{521}
 -i\ep} \,
\frac{1}{ x_3^--x_1^-  - D_{631}  
 -i\ep } \,\frac{1}{ x_3^--x_1^-  - D_{6421}
  -i\ep } \, .
\label{eq:path-form-example-5plus}
\eea
We now go on to describe how coordinate results of this type can be extended to more general theories
and to physical applications.

\section{Applications}
\label{sec:extensions}

In this section, we sketch the relation of the results for the Green functions of massless scalar fields 
to scattering and  to more general massless theories involving fields with spin.   We illustrate how
ultraviolet and infrared divergences appear, and briefly discuss the use of dimensional regularization.

\subsection{Scattering and current-induced configurations}

The path expressions for coordinate space Green functions are, of course, related to perturbative S-matrix elements by Fourier transforms and LSZ reduction \cite{Borinsky:2022msp}.  We will not follow this route in detail, but we can make a useful observation on the basis of the general expression, Eq.\ (\ref{eq:1st-result-massless}).  Contributions to the scattering matrix will require transforms on outgoing external vertices at positions $x_b^\mu$ and incoming at $x_a^\mu$, of the form
\bea
\int d^4x_a \; e^{-ip_a\cdot x_a}\ \int d^4x_b \; e^{ip_b\cdot x_b}\ G^{(\C)}_{\P}\left (\{x_b\}_{\rm out},\{x_a\}_{\rm in}\right )
\ \propto\ \theta(p_a^+)\, \theta(p_b^+)\, .
\label{eq:S-matrix-thetas}
\eea
The theta functions on the right follow from the explicit dependence of the denominators in Eq.\ (\ref{eq:1st-result-massless}) on the minus components of the external vertices, $x_a^-$ and $x_b^-$, whose poles are in the lower half-plane and upper half-plane, respectively.  Equation (\ref{eq:S-matrix-thetas}) confirms the intuitive expectation that a scattering process like $p_1+p_2 \to p_3+ \dots p_n$ is found entirely from the Fourier transforms of diagrams in which the conjugate vertices $x_1$ and $x_2$ are incoming and $x_3 \dots x_n$ are outgoing.

Of particular interest are processes which, as in leptonic annihilation reactions, are initiated by currents.   In QCD, these would be electroweak vector currents, but for our purposes, we can take them to be scalar as well.   In this case, energy will flow into the corresponding coordinate diagrams at a unique earliest, incoming vertex, $\zeta$, with all other external vertices outgoing.    We will see this configuration emerge below in our discussion of cross sections.

\subsection{Theories with spin: eikonal limits and Wilson lines}

So far, our discussion has been entirely in terms of scalar fields.   Naturally, we are interested primarily in applications to gauge and other theories involving fields with spin.     In coordinate space, fermion propagators and triple-vector vertices involve derivatives with respect to the integration variables $y_i^\mu$ that act on scalar propagators, resulting in factors of the general form $z^\mu / (-z^2+i\ep)^2$.     We can, however, extricate the derivative action from the minus integrals that lead to the basic results for scalars, Eqs.\ (\ref{eq:GOtoGCO}) and (\ref{eq:1st-result-massless}).   
To do so, we introduce a lightlike vector, for any choice of plus and transverse components,
\bea
\hat w^\mu \ =\ \left( w^+ \, , \, \hat w^- \equiv \frac{w_{\perp}^2}{2w^+}\, , \, w_{\perp} \right)\, ,  \quad\ \quad \hat w^2=0 \ .
\label{eq:hatw-def}
\eea
With this definition in hand, we can readily check the identity,
\bea
\frac{w^\mu}{(-w^2\ +\ i\ep)^2}  & = & \frac{1}{2w^+}\frac{\partial}{\partial \hat w^-}\ \left[  \frac{-\ \hat w^{\mu}} {-2w^+\left(w^-\ -\ \hat w^- \right) + \, i\ep} \right]\ .
\label{eq:hatw-deriv}
\eea
Using this relation, gauge theory results in coordinate light-cone perturbation theory can be generated directly from scalar expressions entirely in terms of derivatives carried out at fixed values of plus and transverse vertex coordinates.  

Of particular interest in gauge theory are Wilson lines, of the general form,
\bea
\Phi_\beta^{(f)} = {\cal P}  \exp \left [  -ig \int_0^\infty d\lambda\, \beta\cdot A^{(f)}(\lambda\beta) \right ]\, ,
\label{eq:Phi-def}
\eea
where  ${\cal P}$ represents path ordering of the gauge fields  $A^\mu$ in matrix representation $f$, and $\beta$ is the velocity vector that defines the path associated with the Wilson line.   A fundamental example 
is the vacuum expectation of two such lines in distinct directions, $\beta$ and $\bar \beta$, which meet at a color singlet vertex, or cusp, in gauge theory, $\langle 0| { \Phi^{(f)}_{\bar \beta}}^\dagger(0)\, \Phi^{(f)}_\beta(0) |0\rangle$. 
   As recoilless sources of gauge fields, Wilson lines can be incorporated into coordinate light-cone formalism by treating the vertices at which they emit or absorb vector particles as external vertices in the sense introduced above, whose locations move along the path of the line, as in Eq. (\ref{eq:Phi-def}).\footnote{In momentum space, Wilson lines can be generated from the eikonal approximation for the emisson of soft radiation by fast-moving charged particles, by neglecting terms quadratic in soft vector momenta compared to linear, $2p\cdot k \gg k^ 2$, by assigning a vector coupling $-igp^\mu$ with each vertex at which the soft vectors are attached, and by cancelling the dimensional scale of the hard momentum, using $p^\mu/p\cdot k = \beta^\mu/\beta\cdot k$, with $\beta$ dimensionless.   A systematic approach beyond eikonal approximation is described in \cite{Laenen:2008gt}.}
   The path formulation of low order examples of  Wilson line amplitudes was carried out in Ref.\ \cite{Erdogan:2017gyf}, where the results were compared to previous studies in Refs.\ \cite{Laenen:2014jga,Laenen:2015jia}, based on causal propagators.

   \subsection{Ultraviolet divergences and dimensional regularization}

\begin{figure}[t]
\centering
\includegraphics[height=3.5cm]{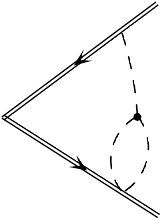}\\
\vspace{1cm}
\caption{Two-loop correction to the cusp, which illustrates ultraviolet singularities.}  
\label{fig:pseudo-self}
\end{figure}

As in any representation of perturbative diagrams, we anticipate both ultraviolet and infrared singularities in the path formalism in four dimensions. The former occur in any configuration
of external vertices, the latter when one or more external vertices are taken to infinity.   To illustrate how ultraviolet singularities emerge in the path formalism, we can consider a two-loop
contribution to the cusp amplitude, shown in Fig.\ \ref{fig:pseudo-self}.   In the figure, the lower (upper) double line represents a Wilson line in light-like direction $\beta$ ($\bar \beta$), with $\beta^2=\bar\beta^2=0$ and $\beta\cdot\bar\beta=1$.
This diagram, in which two scalar lines connect to one of the Wilson lines, occurs in the analysis of two-loop gauge theory calculations in coordinate space \cite{Korchemskaya:1992je,Erdogan:2011yc}.  
We take the upper vertex to be at point $\sigma\bar\beta$, and eventually integrate over $\sigma$. 
Correspondingly, the lower vertex is set at separation $\lambda\beta^\mu$ from the cusp, which is taken at the origin.    The three-point vertex is
at an arbitrary internal point, $y^\mu$, over which we shall also integrate, here in path formalism.   For simplicity, we choose the lower Wilson line
to be in the plus coordinate direction, and the upper in the minus direction.   Increasing plus coordinates are then downward in the figure.   Treating the vertices
on the Wilson lines as external, this diagram is characterized by two paths, both beginnning at the vertex at $\sigma\bar\beta$, which has zero
plus coordinate, and both ending at $\lambda\beta$, with plus coordinate $\lambda$.  There is only a single possible ordering, with $0 \le y^+\le \lambda$.
In terms of the path formalism, with $y^\mu$ the only internal vertex, this is a one-loop diagram.

Following the general form given in Eq.\ (\ref{eq:1st-result-massless}), after doing the $y^-$ integral exactly as outlined above, we find
(including a conventional coupling dependence),
 \bea
G\left (\{\lambda\beta\}_{\rm out},\{\sigma\bar\beta\}_{\rm in}\right ) \ &=&\
   \frac{g^4}{(2\pi)^5}  \int dy^+ \,  \frac{\theta(y^+)}{2y^+}\, \frac{\theta(\lambda-y^+)}{4(\lambda-y^+)^2}
  \nn\\[2mm]
  &\ & \hspace{-15mm}  \times\
\int  d^2y_\perp\,   \frac{(-1)^2}{ \left[  - \frac{y_\perp^2}{2y^+} - \frac{y_\perp^2}{2(\lambda-y^+)} -\sigma -i\ep \right ]^2 } 
\nn\\[2mm]
&\ &  \hspace{-30mm} =
   \frac{1}{8} \left(\frac{g^2}{4\pi^2}\right)^2  \frac{1}{\lambda\sigma}\;  \int_0^\lambda \frac{dy^+}{\lambda-y^+} 
\, ,
 \label{eq:1st-cusp-uv}
\eea
 where in the second equality we have formally done the two-dimensional $y_\perp$ integral, resulting in  a divergent expression.   The singularity at 
 $y^+=\lambda$ corresponds precisely to shrinking the scalar self-energy subdiagram to a point. 
 Note in addition, that when we integrate over the Wilson line positions, $\lambda$ and $\sigma$,
 we encounter additional singularities at the origin, $\lambda,\sigma=0$, which have an ultraviolet interpretation \cite{Erdogan:2014gha,Salas-Bernardez:2022cuw} and are in need of regularization. 
 
To deal with such ultraviolet singularities, we can use dimensional regularization.   Compared 
to our four-dimensional analysis above, which began with the representation
(\ref{eq:amp-rep}), 
in $4-2\vep$ dimensions we can write for the causal propagator (in this case for $z^+>0$),
\bea
\frac{\Gamma(1-\vep)}{ \left [-z^2 + i\ep \right ]^{1-\vep}} 
&=& \frac{-ie^{i\pi\vep/2}}{(2z^+)^{1-\vep}}\, \int_0^{\infty} dE^+\, (E^+)^{-\vep}
 \exp \left [-iE^+ \left( z^- - \frac{z_\perp^2+i\ep}{2z^+} \right ) \right ]
\, .
\nn\\[2mm]
&\ & \hspace{-20mm} = \frac{-i}{2z^+}\, \frac{1}{\Gamma(\vep)}\, \int_0^{\infty} dz'{}^2\, (z'{}^2)^{\vep-1} \, 
\int_0^\infty dE^+\, \exp \left [-iE^+ \left( z^- - \frac{z_\perp^2+z'{}^2 + i\ep}{2z^+} \right ) \right ]\, .
\nn\\
\label{eq:amp-rep-dimreg}
\eea
In the first representation, the factor 
$(E^+)^{-\vep}$ is an obstacle to our integration procedure. This is because a line light-cone energy
$E^+$ is in general a linear combination of path light-cone energies, so that the integrations over the latter would not factorize
the way they did in Eq.\ (\ref{eq:E-integrals-positive-path}).
In the second form, however, the introduction of the $z'{}^2$ integration 
returns all $E^+$ dependence to the exponent, restoring factorization.
The procedure leading to path denominators then
continues exactly as above, with the variable 
${z'}{}^2$ serving to regulate ultraviolet divergences,
like the one in Eq.\ (\ref{eq:1st-cusp-uv}).  Although the resulting
expressions involve extra $z'{}^2$ integrals, this complication is somewhat alleviated
by the factor of $1/\Gamma(\vep)$ for each $z'{}^2$.   We will not implement
such an ultraviolet regularization in this review because our interest is primarily in
how amplitudes, and cross sections below, are built up over finite distances in space-time.

We close this discussion by reemphasizing that the path results developed here,
centered around Eq.\ (\ref{eq:1st-result-massless}), are limited to Green functions with
massless fields.   A connection to massive fields and their propagators in coordinate
space has been given in Ref. \cite{Erdogan:2017gyf}, in terms of dispersion relations.\footnote{We note that
in \cite{Erdogan:2017gyf} it was observed that dimensional regularization can also be implemented
through dispersion integrals.   The representation of the propagator in Eq.\ (\ref{eq:amp-rep-dimreg}) 
is equivalent to the dispersive treatment outlined in Ref. \cite{Erdogan:2017gyf}.}

\section{Cross Sections in Coordinate Representation}
\label{sec3}

In the following, we describe an extension of the analysis of Green functions given above to cross sections.   
To anticipate, we will derive a path description of inclusive cross sections that is very closely related to
the one for the amplitude, Eq.\ (\ref{eq:1st-result-massless}).   The results we present also have a close
correspondence to the flow-ordered expressions for cut diagrams in Refs.\ \cite{Salas-Bernardez:2023aqt} and \cite{Salas-Bernardez:2023zzv}.

For simplicity, we shall focus in this section on inclusive, current-induced processes, taking the vacuum as the initial state.
  The details of an
analysis of weighted cross sections will be presented elsewhere.
Our discussion will complement momentum-space analyses
that reformulate the perturbative calculation of infrared safe cross sections
without the need of constructing infrared-regulated amplitudes and phase space integrals \cite{Soper:1999xk} -- \cite{Ramirez-Uribe:2024rjg}.
Ideally, infrared cancellation is automatic in such calculations, providing, among other things,
integrands that can be computed numerically.
The coordinate approach presented here gives an additional perspective on 
these mechanisms of cancellation.

\subsection{Momentum-inclusive cross sections as coordinate integrals}

We begin with a reformulation of a total cross section in coordinate space, by a simple exercise in Fourier transforms.
We consider  the scalar analog of an  electroweak annihilation cross section in QCD, mediated by a (scalar) current $J(0)$,
to a final state, $F$, with $n$ particles, integrating over all phase space momenta subject only to overall momentum conservation, with total momentum, $q$, $q^2=Q^2$,
\bea
\sigma_F(Q)\ =\ \int \prod_{i=1}^n \frac{d^4p_i}{(2\pi)^4}\ 2\pi \delta_+ (p_i^2) |A(p_1, \dots p_n) |^2  (2\pi)^4\delta^4 \left ( q - \sum_{j=1}^n p_j \right)\, .
\eea
We will refer to such a cross section as ``momentum-inclusive" below.
The amplitude, $A$, that mediates this  process is the matrix element
\bea
A(p_1, \dots p_n) = \langle p_1, \dots p_n | J(0) | 0 \rangle\, ,
\eea
and we assume it can be calculated in perturbation theory.
We denote for the amplitude's Fourier transform
\bea
\tilde A(y_1,\dots y_n,0) \ \equiv \ \prod_{j=1}^n \int \frac{d^4p_i}{(2\pi)^4}\, e^{-ip_j\cdot y_j} A(p_1, \dots p_n)\, ,
\eea
where the current is set at the origin.   Expressing the momentum conservation delta function as an integral over
coordinate vector $\zeta^\mu$, and  using translation invariance on the amplitude, we arrive at 
\bea
\sigma_{F}(Q)\ &=& \int d^4\zeta\, e^{-iq\cdot \zeta}\ \prod_{i=1}^{n} \int d^4w_i \,d^4y_i \,\tilde A^*(w_1,\dots w_n,0)
\nn\\
&\ & \times \prod_{j=1}^{n} \Delta_c\left( w_j-y_j \right )\, \tilde A(y_1,\dots y_n,\zeta)\, ,
\label{eq:sigma-coordinate}
\eea
where $\Delta_c(z)$ is the Fourier transform of the final state on-shell propagator,
\bea
\Delta_c(z) &=& \int \frac{d^4p}{(2\pi)^4} \, e^{-ip\cdot z} (2\pi)\, \delta_+(p^2)
\nn\\[2mm]
&=& \frac{1}{4\pi^2}\, \frac{1}{ -z^2 + i\ep z^0}
\nn\\[2mm]
&=& \frac{1}{4\pi^2}\, \frac{1}{ -2(z^+-i\ep)(z^--i\ep) + z_\perp^2 }\, .
\label{eq:cut-line-coordinate}
\eea
In the final relation, we reexpress the result in light-cone coordinates, which will of course be useful below.
This coordinate representation of the cut propagator is similar to the expression for the causal propagator, Eq.\
(\ref{eq:causal-propagator}), on which the perturbative expansion of the amplitude $\tilde{A}(y_1,\dots y_n)$ is
built, differing only in the infinitesimal imaginary part, which keeps the singularities in both of its light-cone coordinates in the upper half-plane.
The complex conjugate of the causal propagator, from which the conjugate amplitude
is constructed, again differs from (\ref{eq:causal-propagator}) only in its infinitesimal imaginary part,
\bea
\Delta^*(z) &=& 
\frac{1}{4\pi^2}\, \frac{1}{ -z^2 - i\ep }\, .
\label{eq:cc-causal-propagator}
\eea
With these similarlities in mind, it is natural to carry out an analysis  for the cross section
based on the steps we followed above for Green functions.

\begin{figure}
\centering
\includegraphics[height=3cm]{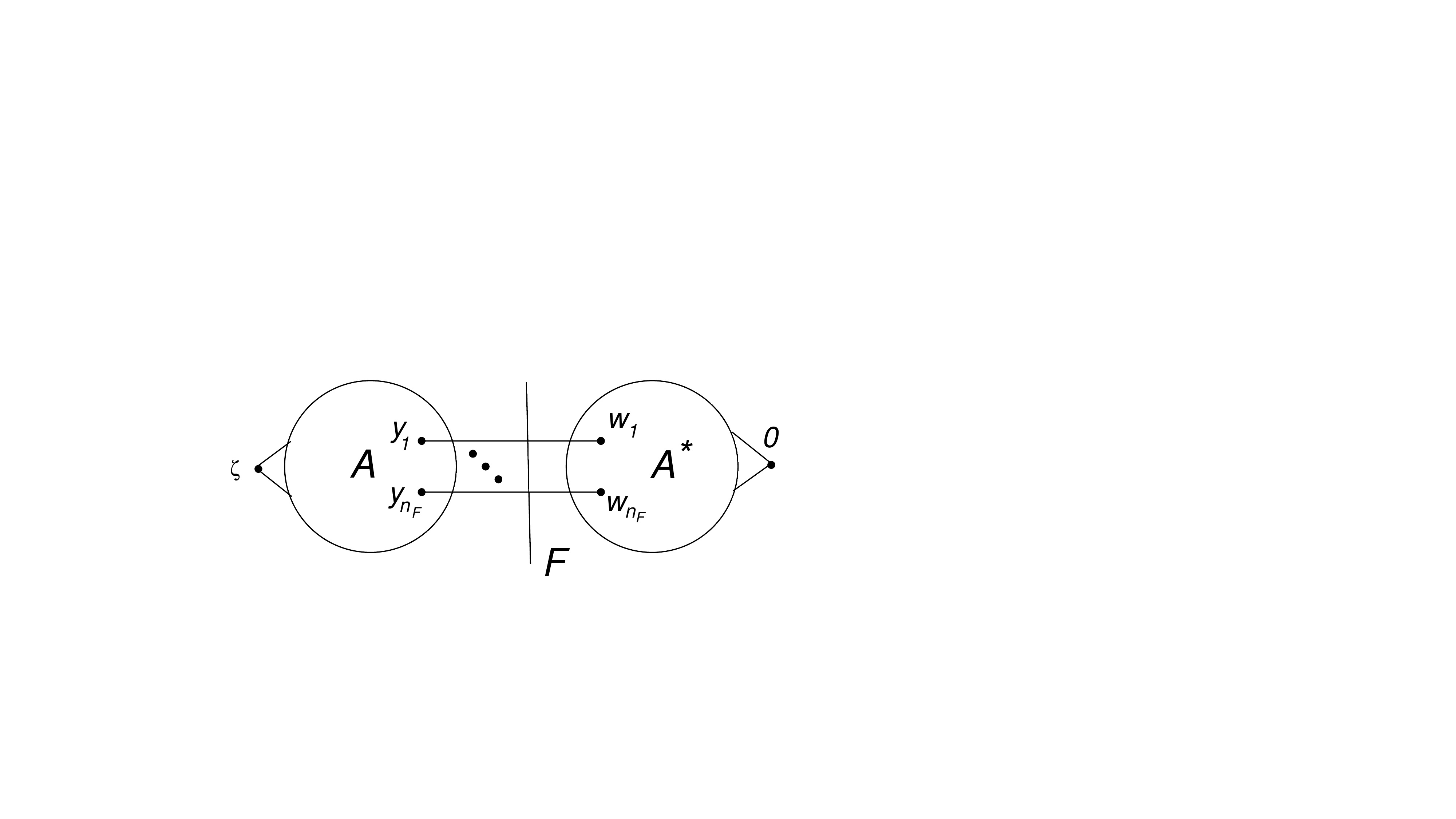}\\
\vspace{1cm}
\caption{Representation of a general cut diagram contribution to Eq.\ (\ref{eq:sigma-coordinate}), which we analyze in coordinate space.}  
\label{fig:sigma-cut}
\end{figure}

\subsection{Path representation for momentum-inclusive cross sections} 

We now set out to analyze an arbitrary graphical contribution to the momentum-inclusive cross section, Eq.\ (\ref{eq:sigma-coordinate}), represented in
Fig.\ \ref{fig:sigma-cut}.   In amplitude $A$, the current at point $\zeta^\mu$ creates a pair of causal propagators, which are connected diagrammatically in $\tilde{A}(y_1,\dots , y_n)$
to a set of $n$ vertices $y_i$, from which $n$ lines emerge into the final state, each propagating  to a point $w_{i'}$
in the complex conjugate amplitude, $\tilde{A}^*(w_1, \dots , w_n)$.  These lines are represented by cut propagators $\Delta_c(w_i-y_i)$.

Generalizing our treatment of Green functions, we start with integral representations for the causal, cut and complex conjugate causal propagators, as functions of the distances between vertices, $z_j$,
\bea
\frac{1}{-z^2_j + i\ep} &=& \frac{-i}{2|z_j^+|}\, \int_{-\infty}^{\infty} dE_j^+\, e^{ -iE_j^+ \left(  z^-_j - \frac{z_{j\perp}^2+i\ep}{2z_j^+} \right ) } \theta\left( E_j^+z_j^+ \right),
\label{eq:causal-rep}
\\[2mm]
\frac{1}{ -2(z_j^+-i\ep)(z_j^--i\ep) + z_{j\perp}^2 }  &=& \frac{-i}{2(z_j^+ - i\ep)} \int_0^\infty dE_j^+ e^{-iE_j^+ \left( z_j^-  - \frac{ z_{j\perp}^2}{2(z_j^+-i\ep)}-i\ep \right ) },
\label{eq:cut-rep}
\\[2mm]
 \frac{1}{ -z^2 - i\ep }  &=& \frac{i}{2|z_j^+|}  \int_{-\infty}^\infty dE_j^+ \,   e^{-iE_j^+ \left ( z_j^- - \frac{z_{j\perp}^2-i\ep}{2z_j^+}  \right )} \theta \left ( - E_j^+z_j^+ \right )\, .
 \nn\\
\label{eq:cc-causal-rep}
\eea
Notice that the representation for the complex conjugate propagator, Eq.\ (\ref{eq:cc-causal-rep}), is related to the representation, (\ref{eq:causal-rep}),  of the causal propagator by an additional change of variables, $E^+_j \rightarrow -E^+_j$.  For the complex conjugate amplitude then, positive light-cone energy flows towards lower plus coordinate values. 

It is useful to organize this contribution to the cross section as a
sum over vertex partial orderings  in the amplitude (${\cal O}_A)$ and in the complex conjugate $(\Ps)$ separately, 
\bea
\sigma_F^{(N_A,N_{\As})}(Q)\ &=&
  \frac{(-g)^{N_A}(g)^{N_{\As}}}{(4\pi^2)^{L_A+L_\As+L_F}}\;  \int d^4\zeta\; e^{-iq\cdot \zeta}\, \sum_\P \;  \prod_{i=1}^{N_A} \int d^2y_{i\perp} \int_{\P} dy_i^+ 
\nn\\[2mm]
&\ & \times  \sum_\Ps\; \prod_{i'=1}^{N_{A^*}} \int d^2w_{i'\perp} \int_{\Ps} dw_{i'}^+ \; G_{\P\Ps} (y_i^+,y_{i\perp},w_{i'}^+,w_{i'\perp},\zeta)\, , \nn\\
\label{eq:sigma-in-orders}
\eea
where we have separated the integrand, $G_{\P\Ps}$, at fixed transverse and plus coordinates.   In the light of our discussion of the amplitude in Sec.\ \ref{sec2}, we anticipate the notation of allowed, that is nonvanishing, orderings of the amplitude ($\P$) and complex conjugate ($\Ps$).  

Once again, as in the reasoning that led to Eq.\ (\ref{eq:E-integrals-positive}) for the amplitude, for each fixed choice of (partial) orderings $\P$ and $\Ps$ of the vertices, we can modify the incidence matrix to adjust the definitions of the propagator arguments $z_j$ so that all $z_j^+$ are positive in the amplitude, and negative in the complex conjugate amplitude.   
Again, we denote the modified incidence matrices as $\eta_{ji}^{(\P)}$ for the amplitude and $\lambda_{j'i'}^{(\Ps)}$ for the complex conjugate.  Notice that at this point, line light-cone energies are not yet conserved at each vertex.

The integrand in Eq.\ (\ref{eq:sigma-in-orders})  is itself still an integral over the minus coordinates of all the internal vertices.  Substituting the integral representations, Eqs.\ (\ref{eq:causal-rep}) - (\ref{eq:cc-causal-rep}) of the $L_A$ lines in the amplitude, $L_{A^*}$ lines in the complex conjugate and $L_F$ lines in the final state, we arrive at 
\bea
G_{\P\Ps} (y_i^+,y_{i\perp},w_{i'}^+,w_{i'\perp},\{x_b\}_{\rm out},\{x_a\}_{\rm in}) &=& i^N\, \prod_{i=1}^{N_A} \int_{-\infty}^\infty dy_i^-\, \prod_{i'=1}^{N_{\As}} \int_{-\infty}^\infty dw_{i'}^-
\nn\\[2mm]
&\ & \hspace{-30mm} \times \prod_{j=1}^{L_A} \frac{-i\theta\left( z_j^+ \right)}{2|z_j|} \int_0^{\infty} dE_j^+\, e^{ -iE_j^+ \left(  z^-_j - \frac{z_j^2+i\ep}{2z_j^+} \right ) } 
\nn\\[2mm]
&\ & \hspace{-30mm} \times
 \prod_{k=1}^{L_F} \frac{-i}{2(z_k^+-i\ep)} \; \int_0^\infty dE_k^+ e^{-iE_k^+ \left( z_k^- -i\ep - \frac{ z_{k\perp}^2}{2(z_j^+-i\ep)} \right ) }
\nn\\[2mm]
&\ & \hspace{-30mm}  \times
\prod_{j{\, }' =1}^{L_{A^*}} \frac{i\theta \left ( - z_{j{\, }' }^+ \right )}{2|z_{j{\, }' }|}\, \int_0^\infty dE_{j{\, }' }^+ \,   e^{-iE_{j{\, }' }^+ \left ( z_{j{\, }' }^- - \frac{z_{j{\, }' \perp}^2 - i\ep}{2z_{j{\, }' }^+}  \right )} \, .
\nn\\
\label{eq:G-plus-ordered}
\eea
The expression on the right is implicitly a function of the vertex plus coordinate orderings, $\P$ and $\Ps$, through the definitions of its variables $z_j^+$.   As indicated by the arguments of $G_{\P\Ps}$ on the left, this form applies for arbitrary choices of incoming and outgoing vertices.
To avoid an unwieldy expression, we provide separately the arguments ($z^\mu$) of propagators in terms of the modified incidence matrices and the positions of  internal vertices of the cut diagram, $y_i^\mu$ and $w_{i'}^\mu$ and the external vertices, which in the general case we label $x_a$ for incoming and $x_b$ for outgoing,   
\bea
z_j &=& \eta^{(\P)}_{ji}y_i + {\eta'}^{(\P)}_{ja}x_a\, ,
\nn\\[2mm]
z_k &=& \lambda^{(\Ps)}_{ki'} w_{i'} + {\lambda'}^{(\Ps)}_{kb}x_b + \eta^{(\P)}_{ki}y_i + {\eta'}^{(\P)}_{ka}x_a\, ,
\nn\\[2mm]
z_{j'} &=& \lambda^{(\Ps)}_{j'i'}w_{i'} + {\lambda'}^{(\Ps)}_{j'b}x_b\, .
\label{eq:z-to-y-w}
\eea
Note that in these equations, index $i$ runs from $1$ to $L_A$ and $i'$ from $1$ to $L_{\As}$.
For coordinate diagrams corresponding to an arbitrary cross section with two particles in the initial state, we have two incoming and two outgoing vertices.
For the current induced diagrams we present below, there is only one $x_a=\zeta$ and one $x_b=0$.

Collecting the exponential factors of the $y_i^-$ from the amplitude and $w_{i'}^-$ from the complex conjugate, we generate, as for the amplitude in Eq.\ (\ref{eq:E-integrals-1}),
 products of light-cone energy-conservation delta functions at each internal vertex.   We find, still with general incoming and outgoing vertices,   
\bea
G_{\P\Ps} (y_i^+,y_{i\perp},w_{i'}^+,w_{i'\perp},\{x_b\}_{\rm out},\{x_a\}_{\rm in}) &=& 
\nn\\[2mm]
&\ & \hspace{-55mm} i^{N-L}\, \times \prod_{j=1}^{L_A} \frac{\theta\left( z_j^+ \right)}{2z^+_j} \int_0^{\infty} dE_j^+\, e^{ -i \sum_{j=1}^{L_A} E_j^+ \left(  {{\eta'}^{(\P)}_{ja}}x_a^- - \frac{z_j^2+i\ep}{2z_j^+} \right ) } 
\nn\\[2mm]
&\ & \hspace{-50mm} \times
 \prod_{k=1}^{L_F} \frac{1}{2(z^+_k-i\ep)} \; \int_0^\infty dE_k^+ e^{-i \sum_{k=1}^{L_F} E_k^+ \left( {\lambda'}^{(\Ps)}_{kb} x_b^- + {\eta'}^{(\P)}_{ka} x_a^-  - \frac{ z_{k\perp}^2}{2(z_k^+-i\ep)}-i\ep \right ) }
\nn\\[2mm]
&\ & \hspace{-50mm}  \times
\prod_{j'=1}^{L_{\As}}\, \frac{\theta \left ( - z_{j'}^+ \right )}{2z^+_{j'}}\, \int_0^\infty dE_{j'}^+ \,   e^{-i\sum_{j'=1}^{L_{\As}} E_{j'}^+ \left ( {\lambda'}^{(\Ps)}_{j'b}x_b^- - \frac{z_{j'\perp}^2-i\ep}{2z_{j'}^+}  \right )} 
\nn\\[2mm]
&\ & \hspace{-50mm}  \times \prod_{i=1}^{N_A}\, (2\pi)\, \delta\left( \sum_{j=1}^{L_A} E_j^+\eta^{(\P)}_{ji} + \sum_{k=1}^{L_{N_F}} E_k^+\eta^{(\P)}_{ki} \right )
\nn\\[2mm]
&\ &  \hspace{-50mm} \times \prod_{i'=1}^{N_{\As}}\, (2\pi)\, \delta\left(  \sum_{k=1}^{L_{N_F}} E_k^+\lambda^{(\Ps)}_{ki'} + \sum_{j'=1}^{L_{A^*}} E_{j'}^+\lambda^{(\Ps)}_{j'i'} \right )
\, ,
\label{eq:G-plus-ordered-deltas}
\eea
where we show explicitly the sums over line indices in the exponents and for each vertex in the arguments of the delta functions.
In Eq.\ (\ref{eq:G-plus-ordered-deltas}), we exhibit the dependence on the minus components of the incoming ($x_a$) and outgoing ($x_b$) vertices in the exponents of the light-cone energy integrals, leaving the transverse and plus dependence as defined by Eq.\ (\ref{eq:z-to-y-w}).   Notice that we have factored an overall $i^{-L}$, using that $z_{j'}^+$ is negative for all lines $j'$ in the conjugate amplitude.

Equation  (\ref{eq:G-plus-ordered-deltas}), is in close correspondence to Eq.\ (\ref{eq:E-integrals-1}) in the final step of the derivation of the path form for an amplitude.   The only differences are in the line $z_j^+$ dependence, which has minus signs in the theta functions  for lines in the complex conjugate amplitude, while 
   there is no relative ordering along the final state lines, $z_k$, for which light-cone energy always flows from amplitude vertices $y_i$ to complex conjugate amplitude vertices $w_i$, regardless of their relative plus component values.

As in Sec.\ \ref{sec2}, the light-cone energy delta functions restrict orders $\P$ and $\Ps$ to those in which light-cone energy is conserved.
Now, again as for the amplitude, we can eliminate as many light-cone energy integrals as there are internal vertices,
by a set of changes of variables, in the form of Eq.\ (\ref{eq:Ej-Ealpha}), in terms of light-cone energies on 
sets of paths.   

For the current-induced process, paths begin at the incoming vertex ($x_a=\zeta$) pass through the amplitude toward
larger $y_i^+$, across the final state (with no plus component order) and through the complex conjugate toward
decreasing $w_{i'}^+$, ending at the outgoing vertex ($x_b=0$), with light-cone energies conserved at each vertex.  
Proceeding as for the amplitude, we can construct sets of paths, $\C$, that cover the full diagram, in terms of whose light-cone energies
 we can express all the line light-cone energies.   Thus, we have
 \bea
 E_l^+\ &=&\ \sum_{{\rm paths}\  \alpha=1}^{L-N} \sigma^{(\C)}_{l\alpha} {\cal E}_\alpha^{({\cal C})}\, ,
 \label{eq:Ej-Ealpha-path}
 \eea
 for any line $z_l$,  in the amplitude, the complex conjugate amplitude and the final state.  Here, $L\equiv L_A+L_{\As}+L_F$ is the
 total number of lines of the cut diagram and $N \equiv N_A+N_{\As}$ the total number of internal vertices.   
 This change of variables, and hence the incidence matrix $\sigma^{(\C)}_{l\alpha}$, can be constructed
 exactly as for the amplitude, 
 following Eqs.\  (\ref{eq:GC-fills}), (\ref{eq:GC-disjoint}) and (\ref{eq:volumes}),
 resulting in an expression for the cross section that is a sum over plus component orderings
 in the amplitude and complex conjugate, and a sum over sets of paths $\C=\{\pi_\alpha\}$ for each ordering.    The full
 integration volume requires, in general, a set of covering sets, $\cal C$, which we denote $\C[\P \Ps]$.
 
 For an example of the procedure, we can consider 
  contributions to a momentum-inclusive cross section
 found by cutting the diagram given in Fig.\ \ref{fig:path-example} between any two of its ordered vertices,
 corresponding to a two- or three-particle final state.   For any such cut, the
 assignments of the path energies specified by the figure is unchanged, while the ordering of vertices by their plus components 
 reverses sign in the cross section, relative to the amplitude, to the right of the cut.

 We can represent each path, $\pi_\alpha$, as the union of amplitude, final state
 and complex  conjugate amplitude segments, in a notation similar to the one for the Green functions,
\bea
 \pi_\alpha\ =\   \pi_\alpha^{(A^*)} \cup z_{\pi_\alpha} \cup \pi_\alpha^{(A)}\, ,
 \label{eq:cut-path-notation}
\eea
where $z_{\pi_\alpha}$ labels the single line from the final state that is on the path.
Because the cut propagators have no ordering, the plus components of the final vertex along any
path in the amplitude and/or  the first vertex encountered by the path in the complex conjugate amplitude,
can go to infinity, and we can think of the paths as extending from a point near the origin ($\zeta^\mu \sim 1/Q$)
out to infinity and back to the origin.
This picture thus fits the closed time nature of Schwinger-Keldysh formalisms \cite{Schwinger:1960qe,Keldysh:1964ud}.

The cross section, Eq.\ (\ref{eq:sigma-in-orders}) is now a sum over allowed orderings, $\P$ and $\Ps$, and covering sets of paths, ${\cal C}$, of  $G^{(\C)}_{\P \Ps}$,
\bea
\sigma_F^{(N_A,N_{\As})}(Q)\ &=& \frac{(-g)^{N_A}(g)^{N_{\As}}}{(2\pi)^{2L-N}}\;  \sum_{\P \Ps} \;  \prod_{i=1}^{N_A} \int d^2y_{i\perp} \int_{\P} dy_i^+ \; 
\nn\\[2mm]
&\ & \times  \prod_{i'=1}^{N_{A^*}} \int d^2w_{i'\perp} \int_{\Ps} dw_{i'}^+ \, \int d^4\zeta\, e^{-iq\cdot \zeta}
\nn\\[2mm]
 &\ & \times 
 \sum_{\C \in {\cal C}[\P \Ps]} G^{(\C)}_{\P \Ps} (y_i^+,y_{i\perp},w_{i'}^+,w_{i'\perp},\zeta) \, .
 \label{eq:sigma-C-o-ostar}
\eea
For each covering set $\cal C$, we can carry out the light-cone energy integrals in Eq.\ (\ref{eq:G-plus-ordered-deltas}) as in the Green functions above. The factors of $i$ cancel and we find familiar products of path denominators,
\bea
G^{(\C)}_{\P \Ps} (y_i^+,y_{i\perp},w_{i'}^+,w_{i'\perp},\zeta) &=&
\prod_{j=1}^{L_A} \frac{\theta(z_j^+)}{2z^+_j} \;   \prod_{k=1}^{L_F} \frac{1}{2(z^+_k-i\ep)} \; \prod_{j'=1}^{L_{A^*}} \frac{\theta(-z_{j'}^+)}{2z^+_{j'}}
 \nn\\[2mm]
&\ & \hspace{-5mm} \times\ \prod_{\pi_\alpha \in {\cal C}}
\frac{-1}{ \left[ -\,  \sum_{l \in L}\, \sigma^{(\C)}_{l\alpha}  \frac{z_{l\perp}^2}{2z_l^+} \, -\,  \zeta^- - i\ep  \right ] }\, ,
\eea
where each path $\pi_\alpha$ in ${\cal C}$ is represented by a denominator determined by its incidence matrix $\sigma^{(\C)}_{l \alpha}$.
Equivalently, in the notation of Eq.\ (\ref{eq:cut-path-notation}) for lines in each path, we have
\bea
G^{(\C)}_{\P \Ps} (y_i^+,y_{i\perp},w_{i'}^+,w_{i'\perp},\zeta) &=& \prod_{j=1}^{L_A} \frac{\theta(z_j^+)}{2z^+_j} \;   \prod_{k=1}^{L_F} \frac{1}{2(z^+_k-i\ep)} \; \prod_{j'=1}^{L_{A^*}} \frac{\theta(-z_{j'}^+)}{2z^+_{j'}}
 \nn\\[2mm]
&\ & \hspace{-35mm} \times\ \prod_{\pi_\alpha \in {\cal C}}
\frac{-1}{ \left[   \sum_{j'\in \pi_\alpha^{(A^*)}} \frac{z_{j'\perp}^2}{2|z_{j'}^+|} -  \frac{z_{\pi_F\perp}^2}{2(z_{\pi_F}^+-i\ep)}
- \sum_{j\in \pi_\alpha^{(A)}} \frac{z_{j\perp}^2}{2z_j^+} - \zeta^- - i\ep  \right ] }\, .
\nn\\
\label{eq:sigma-full-paths}
\eea
This expression is a direct generalization of the form for Green functions, Eq.\ (\ref{eq:1st-result-massless}).   The full contribution of the diagram in question to the momentum-inclusive cross section, Eq.\ (\ref{eq:sigma-C-o-ostar}), is the sum over the remaining integrals of each such term.

For fixed plus coordinates, the paths in Eq.\ (\ref{eq:sigma-full-paths}) extend from
initial point $\zeta$ to some largest vertex plus position, in general different for each path, and back to the origin.   As the plus coordinates
are integrated, these paths extend to future infinity in the $y_i^+$, and for a given final state with $n$ particles, can generate infrared singularities.
We close this section with a few comments on how these singularities occur, how they cancel when we sum over final states, and how
this formalism can be generalized.

\begin{figure}[h]
\centering
\includegraphics[height=6cm]{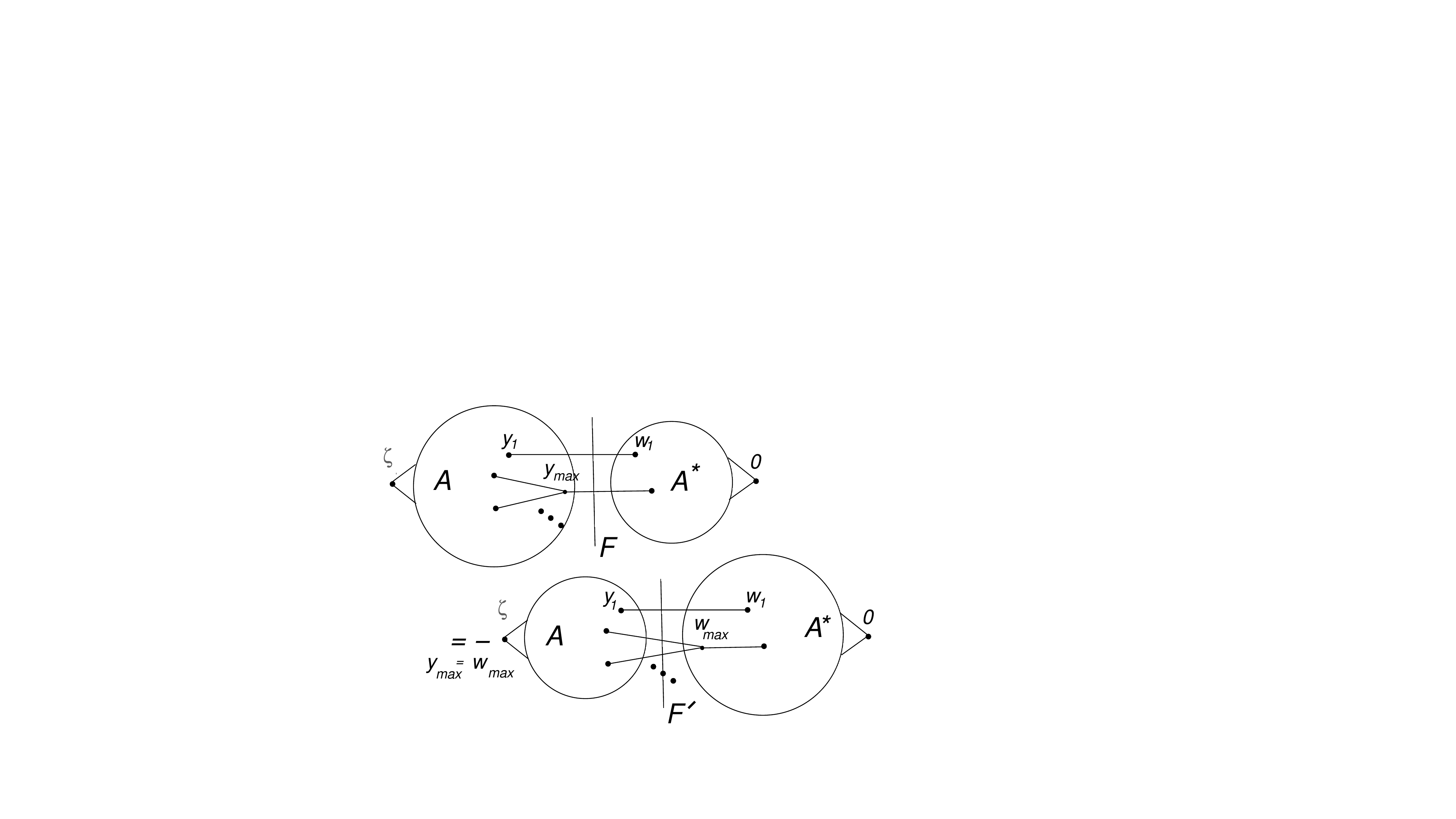}
\caption{Largest time relation.  When the vertex with largest $y^+=y_{\rm max}$ is transferred from the amplitude to the complex conjugate (labeled $w_{\rm max}$), the only difference is a sign, and the sum cancels.}   
\label{fig:cut-diagrams}
\end{figure}

\subsection{IR singularities, cancellation and weighted cross sections}

When all the line coordinates, $z^\mu_j$, in a given path, $\alpha$, in Eq.\ (\ref{eq:sigma-full-paths}) are either bounded or become collinear, the path denominator remains finite even when
the individual components $z_j^\mu$ of the collinear lines become arbitrarily large.   The feature of finiteness in the collinear limit can be seen from the identity,
\bea
 \sum_{j'\in \pi_\alpha^{(A^*)}} \frac{z_{j'\perp}^2}{2|z_{j'}^+|} -  \frac{z_{\pi_F\perp}^2}{2(z_{\pi_F}^+-i\ep)}
- \sum_{j\in \pi_\alpha^{(A)}} \frac{z_{j\perp}^2}{2z_j^+}
=  - \beta^2\, \left(  \sum_{j'\in \pi_\alpha^{(A^*)}} z_{j'}^+ +  z_{\pi_F}^+  + \sum_{j\in \pi_\alpha^{(A)}} z_j^+ \right )\, ,
\nn\\
\eea 
which holds if we demand that $\beta^2\equiv{z_{j\perp}^2}/(2(z_j^+)^2)$ is the same for all the lines, $j$, and we use that the
$z_j^+$ are all positive and the $z_{j'}^+$ are all negative.  Although the individual 
plus components of the lines are large, if they are all collinear their sum must be no larger than $\zeta^+\sim 1/Q$ even as one or more
plus components go to infinity.  This is the origin of collinear singularities.   We can for example, 
imagine coordinates $y^+_{\rm max}$ or $w_{\rm max}^+$  becoming arbitrarily large in Fig.\ \ref{fig:cut-diagrams}.
Correspondingly, the coordinate analog of soft singularities, generally in gauge theories, requires that a set of coordinates
grow without bound in all components.  A general diagram with a fixed final state has both types of singularities,
reflecting the need for infrared regularization in the S-matrix.    

 Collinear singularities cancel point by point when we sum
over final states, $F$.   When we add the integrands of two cross sections when those integrands differ only by moving the 
vertex with the largest plus value from the amplitude to the complex conjugate, they cancel identically,
simply because  the factor $(-g)^{N_A}$ in Eq.\ (\ref{eq:sigma-full-paths}) changes sign, while all path denominators remain
 the same.   
There is no other difference between these terms.   
Such a relation is illustrated in Fig.\ \ref{fig:cut-diagrams}.   

This same cancellation, the ``largest time equation", is the basis of a classic
proof of unitarity \cite{Veltman:1963th,tHooft:1973wag}, and the relation to infrared cancellation in coordinate space was noted in \cite{Sterman:2016zby}.
Here, in fact, the largest time equation implies that {\it all} interaction vertices cancel when they have plus components that are outside the range of $\zeta^+$ to 0.  
 At large $Q^2$, then, the plus components of non-cancelling vertices are squeezed into a segment that is typically of order $1/Q$.  
 The transverse components of these vertices can be large, however, and can naturally be thought of as leading to power corrections as specified by the operator product expansion \cite{Mueller:1984vh}.

Finally,  we note that the path formalism that led to Eq.\ (\ref{eq:sigma-full-paths}) can readily be
extended from fully momentum-inclusive cross sections to a much larger class.   For example, we may consider the cross section for final state $F$ with $n$ particles,
as above, but now
weighted by an arbitrary function of the momenta of the lines in state $F$, $h(p_1, \dots , p_n)$,
\bea
\sigma_F[Q,h] =\ \int \prod_{i=1}^n \frac{d^4p_i}{(2\pi)^4}\ 2\pi \delta_+ (p_i^2)\, h(p_1, \dots , p_n)\, |A(p_1, \dots p_n) |^2  (2\pi)^4\delta^4 \left ( q - \sum_{j=1}^n p_j \right)\, .
\nn\\
\label{eq:sigma-weight}
\eea
The coordinate space analog of the momentum inclusive cross section, Eq.\ (\ref{eq:sigma-coordinate}) is a convolution in phase space, in terms of the 
Fourier transform $\tilde h$ of the weight function,
\bea
\sigma_{F}[Q,h]\ &=& \int d^4\zeta\, e^{-iq\cdot \zeta}\ \prod_{i=1}^{n} \int d^4w_i \,d^4y_i \, \tilde A^*(w_1,\dots w_n,0)
\nn\\
&\ & \hspace{-10mm} \times \prod_{j=1}^{n}\int d^4r_j\,  \Delta_c\left( w_j-y_j - r_j\right )\, \tilde h(r_1,\dots,r_n)\, \tilde A(y_1,\dots y_n,\zeta)\, .
\label{eq:sigma-weight-coordinate}
\eea
Infrared safe weighted and other jet cross sections \cite{Larkoski:2017jix} as well as
energy-energy correlations \cite{Basham:1978bw} -- \cite{Lee:2024esz}
can in principle be treated in this fashion.  

In fact, the path interpretation depends only on the general feature that the
 cut propagator, $\Delta_c(z)$, Eqs.\ (\ref{eq:cut-line-coordinate}) and (\ref{eq:cut-rep}), is the inverse Fourier transform to $z^-$ of a function with only positive frequencies.
 This is the same as the requirement that the cut propagator be analytic in the $z^-$ lower half plane.   
Functions where the usual cut propagator  is replaced by another integral representation
in which the $z^-$ dependence is entirely exponential over the same range will give similar results,
but in general with different path denominators.  
For any such observables, the resulting expressions will be very much in the spirit of Schwinger's method to calculate expectation values of operators
while bypassing the direct computation of transition amplitudes for intermediate states \cite{Schwinger:1960qe}.

\section{Summary}
\label{sec:summary}

We have reviewed the derivation of the coordinate version of light-cone-ordered perturbation theory
for amplitudes of massless fields, and have extended the formalism  to current-induced cross sections 
with specified particle content.  We have seen in both cases that coordinate paths
play a role similar to the role of intermediate states in LCOPT, and that 
for cross sections these paths are closed in time.    We have indicated that the path
formalism can be applied to more general sums over final states, and look forward to further
developments using this perspective on infrared safe and other observables in high energy collisions.

\section*{Acknowledgements}

This work was supported in part by the National Science Foundation, award PHY-2210533.   GS acknowledges the
Simons Center for Geometry and Physics, for enabling conversations at the workshop Energy Operators in Particle Physics,
Quantum Field Theory and Gravity, which were helpful in preparing this review.


\begin{thebibliography}{}

\bibitem{Brodsky:1997de}
S.~J.~Brodsky, H.~C.~Pauli and S.~S.~Pinsky,
``Quantum chromodynamics and other field theories on the light cone,''
Phys. Rept. \textbf{301}, 299-486 (1998)
doi:10.1016/S0370-1573(97)00089-6
[arXiv:hep-ph/9705477 [hep-ph]].

\bibitem{Bardakci:1968zqb}
K.~Bardakci and M.~B.~Halpern,
``Theories at infinite momentum,''
Phys. Rev. \textbf{176}, 1686-1699 (1968)
doi:10.1103/PhysRev.176.1686

\bibitem{Chang:1968bh}
S.~J.~Chang and S.~K.~Ma,
``Feynman rules and quantum electrodynamics at infinite momentum,''
Phys. Rev. \textbf{180}, 1506-1513 (1969)
doi:10.1103/PhysRev.180.1506

\bibitem{Kogut:1969xa}
J.~B.~Kogut and D.~E.~Soper,
``Quantum Electrodynamics in the Infinite Momentum Frame,''
Phys. Rev. D \textbf{1}, 2901-2913 (1970)
doi:10.1103/PhysRevD.1.2901

\bibitem{Weinberg:1966jm}
S.~Weinberg,
``Dynamics at infinite momentum,''
Phys. Rev. \textbf{150}, 1313-1318 (1966)
doi:10.1103/PhysRev.150.1313

\bibitem{Bjorken:1970ah}
J.~D.~Bjorken, J.~B.~Kogut and D.~E.~Soper,
``Quantum Electrodynamics at Infinite Momentum: Scattering from an External Field,''
Phys. Rev. D \textbf{3}, 1382 (1971)
doi:10.1103/PhysRevD.3.1382

\bibitem{Lepage:1980fj}
G.~P.~Lepage and S.~J.~Brodsky,
``Exclusive Processes in Perturbative Quantum Chromodynamics,''
Phys. Rev. D \textbf{22}, 2157 (1980)
doi:10.1103/PhysRevD.22.2157

\bibitem{Sterman:1978bj}
G.~F.~Sterman,
``Mass Divergences in Annihilation Processes. 2. Cancellation of Divergences in Cut Vacuum Polarization Diagrams,''
Phys. Rev. D \textbf{17}, 2789 (1978)
doi:10.1103/PhysRevD.17.2789

\bibitem{Sterman:1979uw}
G.~F.~Sterman,
``Zero Mass Limit for a Class of Jet Related Cross-sections,''
Phys. Rev. D \textbf{19}, 3135 (1979)
doi:10.1103/PhysRevD.19.3135

\bibitem{Erdogan:2017gyf}
O.~Erdo\u{g}an and G.~Sterman,
``Path description of coordinate-space amplitudes,''
Phys. Rev. D \textbf{95}, no.11, 116015 (2017)
doi:10.1103/PhysRevD.95.116015
[arXiv:1705.04539 [hep-th]].

\bibitem{Veltman:1963th}
M.~J.~G.~Veltman,
``Unitarity and causality in a renormalizable field theory with unstable particles,''
Physica \textbf{29}, 186-207 (1963)
doi:10.1016/S0031-8914(63)80277-3

\bibitem{tHooft:1973wag}
G.~'t Hooft and M.~J.~G.~Veltman,
``DIAGRAMMAR,''
NATO Sci. Ser. B \textbf{4}, 177-322 (1974)
doi:10.1007/978-1-4684-2826-1\_5

\bibitem{Baier:1996sk}
R.~Baier, Y.~L.~Dokshitzer, A.~H.~Mueller, S.~Peigne and D.~Schiff,
``Radiative energy loss and p(T) broadening of high-energy partons in nuclei,''
Nucl. Phys. B \textbf{484}, 265-282 (1997)
doi:10.1016/S0550-3213(96)00581-0
[arXiv:hep-ph/9608322 [hep-ph]].

\bibitem{Zakharov:1997uu}
B.~G.~Zakharov,
``Radiative energy loss of high-energy quarks in finite size nuclear matter and quark - gluon plasma,''
JETP Lett. \textbf{65}, 615-620 (1997)
doi:10.1134/1.567389
[arXiv:hep-ph/9704255 [hep-ph]].

\bibitem{Zakharov:1998sv}
B.~G.~Zakharov,
``Light cone path integral approach to the Landau-Pomeranchuk-Migdal effect,''
Phys. Atom. Nucl. \textbf{61}, 838-854 (1998)
[arXiv:hep-ph/9807540 [hep-ph]].

\bibitem{Arnold:2023qwi}
P.~Arnold, O.~Elgedawy and S.~Iqbal,
``Landau-Pomeranchuk-Migdal effect in sequential bremsstrahlung: Gluon shower development,''
Phys. Rev. D \textbf{108}, no.7, 074015 (2023)
doi:10.1103/PhysRevD.108.074015
[arXiv:2302.10215 [hep-ph]].

\bibitem{Caucal:2023fsf}
P.~Caucal, F.~Salazar, B.~Schenke, T.~Stebel and R.~Venugopalan,
``Back-to-Back Inclusive Dijets in Deep Inelastic Scattering at Small x: Complete NLO Results and Predictions,''
Phys. Rev. Lett. \textbf{132}, no.8, 081902 (2024)
doi:10.1103/PhysRevLett.132.081902
[arXiv:2308.00022 [hep-ph]].

\bibitem{Balitsky:1995ub}
I.~Balitsky,
``Operator expansion for high-energy scattering,''
Nucl. Phys. B \textbf{463}, 99-160 (1996)
doi:10.1016/0550-3213(95)00638-9
[arXiv:hep-ph/9509348 [hep-ph]].

\bibitem{Kovchegov:1999yj}
Y.~V.~Kovchegov,
``Small x F(2) structure function of a nucleus including multiple pomeron exchanges,''
Phys. Rev. D \textbf{60}, 034008 (1999)
doi:10.1103/PhysRevD.60.034008
[arXiv:hep-ph/9901281 [hep-ph]].

\bibitem{Gelis:2010nm}
F.~Gelis, E.~Iancu, J.~Jalilian-Marian and R.~Venugopalan,
``The Color Glass Condensate,''
Ann. Rev. Nucl. Part. Sci. \textbf{60}, 463-489 (2010)
doi:10.1146/annurev.nucl.010909.083629
[arXiv:1002.0333 [hep-ph]].

\bibitem{Borinsky:2022msp}
M.~Borinsky, Z.~Capatti, E.~Laenen and A.~Salas-Bern\'ardez,
``Flow-oriented perturbation theory,''
JHEP \textbf{01}, 172 (2023)
doi:10.1007/JHEP01(2023)172
[arXiv:2210.05532 [hep-th]].

\bibitem{Salas-Bernardez:2023aqt}
A.~Salas-Bern\'ardez, M.~Borinsky, Z.~Capatti and E.~Laenen,
``Flow Oriented Perturbation Theory,''
PoS \textbf{RADCOR2023}, 026 (2024)
doi:10.22323/1.432.0026
[arXiv:2310.09708 [hep-th]].

\bibitem{Salas-Bernardez:2023zzv}
A.~Salas-Bern\'ardez,
``Analytical Quantum Field methods in Particle Physics,''
[arXiv:2309.12707 [hep-ph]].

\bibitem{Capatti:2022mly}
Z.~Capatti,
``Exposing the threshold structure of loop integrals,''
Phys. Rev. D \textbf{107}, no.5, L051902 (2023)
doi:10.1103/PhysRevD.107.L051902
[arXiv:2211.09653 [hep-th]].

\bibitem{Sterman:2023xdj}
G.~Sterman and A.~Venkata,
``Local infrared safety in time-ordered perturbation theory,''
JHEP \textbf{02}, 101 (2024)
doi:10.1007/JHEP02(2024)101
[arXiv:2309.13023 [hep-ph]].

\bibitem{Yan:1973qg}
T.~M.~Yan,
``Quantum field theories in the infinite momentum frame. 4. Scattering matrix of vector and Dirac fields and perturbation theory,''
Phys. Rev. D \textbf{7}, 1780-1800 (1973)
doi:10.1103/PhysRevD.7.1780

\bibitem{Collins:2018aqt}
J.~Collins,
``The non-triviality of the vacuum in light-front quantization: An elementary treatment,''
[arXiv:1801.03960 [hep-ph]].

\bibitem{Laenen:2008gt}
E.~Laenen, G.~Stavenga and C.~D.~White,
``Path integral approach to eikonal and next-to-eikonal exponentiation,''
JHEP \textbf{03}, 054 (2009)
doi:10.1088/1126-6708/2009/03/054
[arXiv:0811.2067 [hep-ph]].

\bibitem{Laenen:2014jga} 
  E.~Laenen, K.~J.~Larsen and R.~Rietkerk,
  ``Imaginary parts and discontinuities of Wilson line correlators,''
  Phys.\ Rev.\ Lett.\  {\bf 114}, no. 18, 181602 (2015)
  doi:10.1103/PhysRevLett.114.181602
  [arXiv:1410.5681 [hep-th]].

\bibitem{Laenen:2015jia} 
  E.~Laenen, K.~J.~Larsen and R.~Rietkerk,
  ``Position-space cuts for Wilson line correlators,''
  JHEP {\bf 1507}, 083 (2015)
  doi:10.1007/JHEP07(2015)083
  [arXiv:1505.02555 [hep-th]].
  
\bibitem{Korchemskaya:1992je}
I.~A.~Korchemskaya and G.~P.~Korchemsky,
``On lightlike Wilson loops,''
Phys. Lett. B \textbf{287}, 169-175 (1992)
doi:10.1016/0370-2693(92)91895-G

\bibitem{Erdogan:2011yc}
O.~Erdo\u{g}an and G.~Sterman,
``Gauge Theory Webs and Surfaces,''
Phys. Rev. D \textbf{91}, no.1, 016003 (2015)
doi:10.1103/PhysRevD.91.016003
[arXiv:1112.4564 [hep-th]].

\bibitem{Erdogan:2014gha}
O.~Erdo\u{g}an and G.~Sterman,
``Ultraviolet divergences and factorization for coordinate-space amplitudes,''
Phys. Rev. D \textbf{91}, no.6, 065033 (2015)
doi:10.1103/PhysRevD.91.065033
[arXiv:1411.4588 [hep-ph]].

\bibitem{Salas-Bernardez:2022cuw}
A.~Salas-Bern\'ardez,
``Explicit computation of jet functions in coordinate space,''
Nucl. Phys. B \textbf{985}, 116024 (2022)
doi:10.1016/j.nuclphysb.2022.116024
[arXiv:2205.05423 [hep-ph]].

\bibitem{Soper:1999xk}
D.~E.~Soper,
``Techniques for QCD calculations by numerical integration,''
Phys. Rev. D \textbf{62}, 014009 (2000)
doi:10.1103/PhysRevD.62.014009
[arXiv:hep-ph/9910292 [hep-ph]].

\bibitem{Becker:2010ng}
S.~Becker, C.~Reuschle and S.~Weinzierl,
``Numerical NLO QCD calculations,''
JHEP \textbf{12}, 013 (2010)
doi:10.1007/JHEP12(2010)013
[arXiv:1010.4187 [hep-ph]].

\bibitem{Gnendiger:2017pys}
C.~Gnendiger, A.~Signer, D.~St\"ockinger, A.~Broggio, A.~L.~Cherchiglia, F.~Driencourt-Mangin, A.~R.~Fazio, B.~Hiller, P.~Mastrolia and T.~Peraro, \textit{et al.}
``To ${d}$, or not to ${d}$: recent developments and comparisons of regularization schemes,''
Eur. Phys. J. C \textbf{77}, no.7, 471 (2017)
doi:10.1140/epjc/s10052-017-5023-2
[arXiv:1705.01827 [hep-ph]].

\bibitem{Anastasiou:2018rib}
C.~Anastasiou and G.~Sterman,
``Removing infrared divergences from two-loop integrals,''
JHEP \textbf{07}, 056 (2019)
doi:10.1007/JHEP07(2019)056
[arXiv:1812.03753 [hep-ph]].

\bibitem{Capatti:2020ytd}
Z.~Capatti, V.~Hirschi, D.~Kermanschah, A.~Pelloni and B.~Ruijl,
``Manifestly Causal Loop-Tree Duality,''
[arXiv:2009.05509 [hep-ph]].

\bibitem{Capatti:2020xjc}
Z.~Capatti, V.~Hirschi, A.~Pelloni and B.~Ruijl,
``Local Unitarity: a representation of differential cross-sections that is locally free of infrared singularities at any order,''
JHEP \textbf{04}, 104 (2021)
doi:10.1007/JHEP04(2021)104
[arXiv:2010.01068 [hep-ph]].

\bibitem{TorresBobadilla:2020ekr}
W.~J.~Torres Bobadilla, G.~F.~R.~Sborlini, P.~Banerjee, S.~Catani, A.~L.~Cherchiglia, L.~Cieri, P.~K.~Dhani, F.~Driencourt-Mangin, T.~Engel and G.~Ferrera, \textit{et al.}
``May the four be with you: Novel IR-subtraction methods to tackle NNLO calculations,''
Eur. Phys. J. C \textbf{81}, no.3, 250 (2021)
doi:10.1140/epjc/s10052-021-08996-y
[arXiv:2012.02567 [hep-ph]].

\bibitem{Kermanschah:2021wbk}
D.~Kermanschah,
``Numerical integration of loop integrals through local cancellation of threshold singularities,''
JHEP \textbf{01}, 151 (2022)
doi:10.1007/JHEP01(2022)151
[arXiv:2110.06869 [hep-ph]].

\bibitem{Rios-Sanchez:2024xtv}
J.~Rios-Sanchez and G.~Sborlini,
``Toward multiloop local renormalization within causal loop-tree duality,''
Phys. Rev. D \textbf{109}, no.12, 125004 (2024)
doi:10.1103/PhysRevD.109.125004
[arXiv:2402.13995 [hep-th]].

\bibitem{Anastasiou:2024xvk}
C.~Anastasiou, J.~Karlen, G.~Sterman and A.~Venkata,
``Locally finite two-loop amplitudes for electroweak production through gluon fusion,''
JHEP \textbf{11}, 043 (2024)
doi:10.1007/JHEP11(2024)043
[arXiv:2403.13712 [hep-ph]].

\bibitem{Ramirez-Uribe:2024rjg}
S.~Ram\'\i{}rez-Uribe, P.~K.~Dhani, G.~F.~R.~Sborlini and G.~Rodrigo,
``Rewording Theoretical Predictions at Colliders with Vacuum Amplitudes,''
Phys. Rev. Lett. \textbf{133}, no.21, 21 (2024)
doi:10.1103/PhysRevLett.133.211901
[arXiv:2404.05491 [hep-ph]].


\bibitem{Schwinger:1960qe}
J.~S.~Schwinger,
``Brownian motion of a quantum oscillator,''
J. Math. Phys. \textbf{2}, 407-432 (1961)
doi:10.1063/1.1703727

\bibitem{Keldysh:1964ud}
L.~V.~Keldysh,
``Diagram technique for nonequilibrium processes,''
Zh. Eksp. Teor. Fiz. \textbf{47}, 1515-1527 (1964)
doi:10.1142/9789811279461\_0007

\bibitem{Sterman:2016zby}
G.~Sterman and O.~Erdogan,
``A coordinate description of partonic processes,''
PoS \textbf{RADCOR2015}, 027 (2016)
doi:10.22323/1.235.0027
[arXiv:1602.00943 [hep-ph]].

\bibitem{Mueller:1984vh}
A.~H.~Mueller,
``On the Structure of Infrared Renormalons in Physical Processes at High-Energies,''
Nucl. Phys. B \textbf{250}, 327-350 (1985)
doi:10.1016/0550-3213(85)90485-7

\bibitem{Larkoski:2017jix}
A.~J.~Larkoski, I.~Moult and B.~Nachman,
``Jet Substructure at the Large Hadron Collider: A Review of Recent Advances in Theory and Machine Learning,''
Phys. Rept. \textbf{841}, 1-63 (2020)
doi:10.1016/j.physrep.2019.11.001
[arXiv:1709.04464 [hep-ph]].

\bibitem{Basham:1978bw}
C.~L.~Basham, L.~S.~Brown, S.~D.~Ellis and S.~T.~Love,
``Energy Correlations in electron - Positron Annihilation: Testing QCD,''
Phys. Rev. Lett. \textbf{41}, 1585 (1978)
doi:10.1103/PhysRevLett.41.1585

\bibitem{Ore:1979ry}
F.~R.~Ore, Jr. and G.~F.~Sterman,
``An operator approached to weighted cross sections,''
Nucl. Phys. B \textbf{165}, 93-118 (1980)
doi:10.1016/0550-3213(80)90308-9

\bibitem{Sveshnikov:1995vi}
N.~A.~Sveshnikov and F.~V.~Tkachov,
``Jets and quantum field theory,''
Phys. Lett. B \textbf{382}, 403-408 (1996)
doi:10.1016/0370-2693(96)00558-8
[arXiv:hep-ph/9512370 [hep-ph]].

\bibitem{Tkachov:1995kk}
F.~V.~Tkachov,
``Measuring multi - jet structure of hadronic energy flow or What is a jet?,''
Int. J. Mod. Phys. A \textbf{12}, 5411-5529 (1997)
doi:10.1142/S0217751X97002899
[arXiv:hep-ph/9601308 [hep-ph]].

\bibitem{Korchemsky:1997sy}
G.~P.~Korchemsky, G.~Oderda and G.~F.~Sterman,
``Power corrections and nonlocal operators,''
AIP Conf. Proc. \textbf{407}, no.1, 988 (1997)
doi:10.1063/1.53732
[arXiv:hep-ph/9708346 [hep-ph]].

\bibitem{Korchemsky:1999kt}
G.~P.~Korchemsky and G.~F.~Sterman,
``Power corrections to event shapes and factorization,''
Nucl. Phys. B \textbf{555}, 335-351 (1999)
doi:10.1016/S0550-3213(99)00308-9
[arXiv:hep-ph/9902341 [hep-ph]].

\bibitem{Bauer:2008dt}
C.~W.~Bauer, S.~P.~Fleming, C.~Lee and G.~F.~Sterman,
``Factorization of e+e- Event Shape Distributions with Hadronic Final States in Soft Collinear Effective Theory,''
Phys. Rev. D \textbf{78}, 034027 (2008)
doi:10.1103/PhysRevD.78.034027
[arXiv:0801.4569 [hep-ph]].

\bibitem{Hofman:2008ar}
D.~M.~Hofman and J.~Maldacena,
``Conformal collider physics: Energy and charge correlations,''
JHEP \textbf{05}, 012 (2008)
doi:10.1088/1126-6708/2008/05/012
[arXiv:0803.1467 [hep-th]].

\bibitem{Belitsky:2013ofa}
A.~V.~Belitsky, S.~Hohenegger, G.~P.~Korchemsky, E.~Sokatchev and A.~Zhiboedov,
``Energy-Energy Correlations in N=4 Supersymmetric Yang-Mills Theory,''
Phys. Rev. Lett. \textbf{112}, no.7, 071601 (2014)
doi:10.1103/PhysRevLett.112.071601
[arXiv:1311.6800 [hep-th]].

\bibitem{Chen:2020vvp}
H.~Chen, I.~Moult, X.~Zhang and H.~X.~Zhu,
``Rethinking jets with energy correlators: Tracks, resummation, and analytic continuation,''
Phys. Rev. D \textbf{102}, no.5, 054012 (2020)
doi:10.1103/PhysRevD.102.054012
[arXiv:2004.11381 [hep-ph]].

\bibitem{Korchemsky:2019nzm}
G.~P.~Korchemsky,
``Energy correlations in the end-point region,''
JHEP \textbf{01}, 008 (2020)
doi:10.1007/JHEP01(2020)008
[arXiv:1905.01444 [hep-th]].

\bibitem{Lee:2022ige}
K.~Lee, B.~Me\c{c}aj and I.~Moult,
``Conformal Colliders Meet the LHC,''
[arXiv:2205.03414 [hep-ph]].

\bibitem{Lee:2024esz}
K.~Lee, A.~Pathak, I.~W.~Stewart and Z.~Sun,
``Nonperturbative Effects in Energy Correlators: From Characterizing Confinement Transition to Improving \ensuremath{\alpha}s Extraction,''
Phys. Rev. Lett. \textbf{133}, no.23, 231902 (2024)
doi:10.1103/PhysRevLett.133.231902
[arXiv:2405.19396 [hep-ph]].


\end{thebibliography}
\end{document}